\documentclass[14pt]{jpp}

\usepackage{color}
\usepackage{graphicx}
\usepackage{natbib}

\ifCUPmtlplainloaded \else
  \checkfont{eurm10}
  \iffontfound
    \IfFileExists{upmath.sty}
      {\typeout{^^JFound AMS Euler Roman fonts on the system,
                   using the 'upmath' package.^^J}%
       \usepackage{upmath}}
      {\typeout{^^JFound AMS Euler Roman fonts on the system, but you
                   dont seem to have the}%
       \typeout{'upmath' package installed. JPP.cls can take advantage
                 of these fonts, if you use 'upmath' package.^^J}%
      }
  \else
  \fi
\fi

\ifCUPmtlplainloaded \else
  \checkfont{msam10}
  \iffontfound
    \IfFileExists{amssymb.sty}
      {\typeout{^^JFound AMS Symbol fonts on the system, using the
                'amssymb' package.^^J}%
       \usepackage{amssymb}%

      }{}
  \fi
\fi

\ifCUPmtlplainloaded \else
  \IfFileExists{amsbsy.sty}
    {\typeout{^^JFound the 'amsbsy' package on the system, using it.^^J}%
     \usepackage{amsbsy}}
    {\providecommand\boldsymbol[1]{\mbox{\boldmath $##1$}}}
\fi

\newcommand\bfB{\boldsymbol{B}}
\newcommand\bfE{\boldsymbol{E}}
 
\newcommand{\bfu}{\mathbf{u}}

\newcommand{\tsP}{\mathbb{P}}

\newsavebox{\astrutbox}
\sbox{\astrutbox}{\rule[-5pt]{0pt}{20pt}}

\newcommand\alfven{Alfv\'en\ }

\title[Separatrices: the crux of reconnection]{Separatrices: the crux of reconnection}

\author[G. Lapenta]%
{Giovanni Lapenta$^1$\thanks{Email address for correspondence: giovanni.lapenta@wis.kuleuven.be}, Stefano Markidis$^2$, Andrey Divin$^{3,4}$, David Newman$^5$, Martin Goldman$^5$}

\affiliation{$^1$Center for mathematical Plasma Astrophysics, Department Wiskunde, University of Leuven, KU Leuven, Belgium\\
$^2$High Performance Computing and Visualization Department, KTH Royal Institute of
Technology, Stockholm, Sweden\\
$^3$Swedish Institute of Space Physics, Uppsala, Sweden\\
$^4$St. Petersburg State University, St. Petersburg, Russia.\\
$^5$University of Colorado, Boulder, USA}

\pubyear{}
\volume{}
\pagerange{}
\date{}
\begin{document}

\maketitle

\begin{abstract}
Reconnection is one of the key processes in astrophysical and laboratory plasmas: it is the opposite of a dynamo. Looking at energy, a dynamo transforms kinetic energy in magnetic energy while reconnection takes magnetic energy and returns is to its kinetic form. Most plasma processes at their core involve first storing magnetic energy accumulated over time and then releasing it suddenly. We focus here on this release. A key concept in analysing reconnection is that of the separatrix, a surface (line in 2D) that separates the fresh unperturbed plasma embedded in magnetic field lines not yet  reconnected with the hotter exhaust embedded in reconnected field lines. In kinetic physics, the separatrices become a layer where many key processes develop. We present here new results relative to the processes at the separatrices that regulate the plasma flow, the energisation of the species, the electromagnetic fields and the instabilities developing at the separatrices.
\end{abstract}


\section{Introduction}
Magnetic reconnection \citep{biskamp, priest-forbes, birn-priest} is a process with two key ingredients. First, the magnetic field lines break their topological connection and form a new connectivity where points that were on different filed lines become connected by the same field line and viceversa. Second, energy stored in magnetic field is released and converted into particle energy, ordered kinetic energy of flows or thermal energy. 
Depending on the point of view more emphasis and attention goes to one or the other aspect above. 

Traditionally, the topological concepts have attracted a considerable attention for their mathematical attractiveness \citep{lau1990three}. The focus on topology has drawn the attention to the loci of field line breakage and reconnection, the x-points in 2D and more generally the null points in 3D. When a full kinetic treatment is used \citep{birnGEM}, there loci are surrounded by regions where the electrons and ions move at speeds differing from each other and from the $\bfE\times\bfB$ drift. These regions are called diffusion regions and typically represented as nested boxes, with the outer pertaining to ions decoupling and the inner to electrons decoupling (e.g. see \citet{birn-priest}).

This overwhelming focus on the x-points and the surrounding diffusion regions has led to a vast literature and the planing and execution of the Magnetospheric Multiscale Mission (MMS)~\citep{curtis1999magnetospheric} that will coronate the last decades of advances in understanding these regions with in situ measurements at resolutions and cadences unparalleled by previous missions.

Nevertheless, the attention is now turning to the energy aspect. From a no-nonsense engineering point of view, energy is what matters. To interpret experiments, astrophysical events and the processes involved in the Sun-Earth connection (also called space weather), energy considerations figure on top of the list of priorities. 

The energy budget and the flow of energy in reconnection has also been studied based on simulation \citep{birn2014forced} and observational data \citep{hamrin2012role,eastwood2013energy}. The attention, however, has typically focused  on the x-point and the surrounding regions of electron and ion diffusion, extended by the recent discovery of elongated electron jets emerging from the electron diffusion region \citep{phan++2001}. Here we focus specifically on the role of the separatrices in the energy transformations linked with reconnection. 

In, 2D the separatrices are magnetic field lines emerging from the x-point and divide lines that have not yet reconnected form lines already reconnected. In 3D, the separatrices become surfaces that similarly divide volumes filled with different classes of lines. Separatrices and their existence are linked with the very existence of topological reconnection \citep{pontin2011three}. A measure of reconnection is based on defining the squashing factor to identify the quasi-separatrix layer, a topological concept extending the concept of separatrix proper \citep{titov2009slip, restante2013geometrical}.

We report here on our recent results relative to the physics developing along the separatrices. 

Section~\ref{sect-flow} focuses on the flow pattern of electrons and ions. The flow is show to be quite different from the naive two nested box model that is useful in understanding many other aspects of reconnection but is of limited use in describing particle flow. A more correct cartoon focuses not he role of the separatrices where much of the flow features are concentrated. The flow is characterised by a strong parallel flow, but also the perpendicular flow plays a key role. At the separatrices, the electrons violate severely the frozen in condition in consequence of the narrow electric field features that cannot be represented by the gyrocenter approximation. 

Section~\ref{sect-hall} addresses the evergreen of kinetic reconnection: the Hall physics. After summarising a possible narrative of the physics of the Hall mechanism based on the correlation between electron motion and field line bending, we review the structure of the Hall signature, observing the electrostatic nature of the Hall electric field. Two very peculiar signatures are observed: bipolar signature in the parallel electric field and string of the perpendicular electric field. These are signs of instabilities caused by the strong flows and are the subject of Sect.~\ref{sect-instabilities}. 

Section~\ref{sect-instabilities} analyses two types of instabilities generated by the strong flows and by the strong shears in these flows. The focus is on streaming instabilities that cause electron holes and on Kelvin-Helmholtz-type instabilities. Each has distinctly different signatures and develop in planes orthogonal to each other. A full 3D study is needed to follow the evolution of both instabilities together, the topic of the following Sect.~/ref{sec:3dlociofinstability}.

The last aspect analysed is that of the energy consequences of the physical processes described in the present paper. The focus is on both types of energisation: the increase in bulk flow energy and in thermal energy. For both the separatrices are key players. The bulk flow energisation is shown to be an integral part of the kinetic Alfven wave wing that forms at the separatrix as part of the Hall physics. The parallel thermal energy, instead, is increased at the separatrix primarily as a consequence of the electrostatic streaming instabilities described in Sect.~\ref{sect-instabilities}. Additional mechanisms for heating are active in the exhaust via the Fermi and adiabatic mechanism.

\section{Separatrices as flow gates}
\label{sect-flow}

To understand the dynamics of kinetic reconnection, concepts typical of MHD reconnection are often used. Two schemes have received great attention.

The first is based on the Sweet-Parker~\citep{Parker72,1958IAUS....6..123S} model. When applied to kinetic reconnection, a central site of reconnection is described as being surrounded by two boxes with elongated aspect ratio. 
\begin{figure}
  \centering
  \includegraphics[width=\columnwidth]{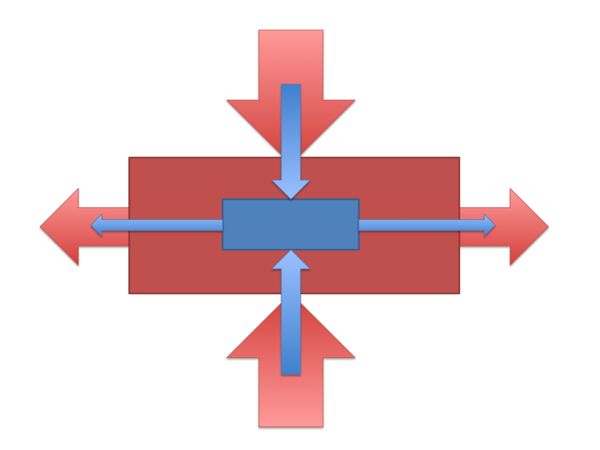}
  \caption{Cartoon of the electron and ion diffusion region in kinetic reconnection. The electrons and ions enter the boxes from the horizontal sides and exit (accelerated from the vertical sides). The outer box is the ion diffusion region where only the ions are demagnetized, while the inner box is the electron diffusion region where the electrons also loose their connection with the magnetic field lines. }
\label{boxes}
\end{figure}
Figure \ref{boxes} represents the concept pictorially. The diagonals of the boxes are the separatrices, that divide the unreconnected from the reconnected plasma.  The outer box has both species entering from the long horizontal sides and exiting from the short vertical sides. This box is identified with the ion diffusion region where the ions no longer move with the same speed as  the field lines and their motion cannot be described by the $\bfE\times\bfB$ drift. In this box, however, the electrons remain still tied to the magnetic field motion.

The inner box is the electron diffusion region, a region where now the electrons loose their ties with the magnetic field motion and with the $\bfE\times\bfB$ drift. The electrons are visualised to enter the long horizontal side and exit the short vertical sides. An electron jet is known to form at the exit of this box~\citep{fujimoto2006time,daughton2006fully} but the presence of guide fields can disrupt its flow~\citep{Goldman:2011, le2013regimes}. The Sweet-Parker-type visualization is useful to obtain scaling properties of the kinetic reconnection process.

The second conceptual model is that of Petschek~\citep{petschek} where  the plasma is not seen as moving in and out  boxes, but rather to traverse standing shocks where energy is exchanged and the plasma species are energized. As we will see below, this description is  more relevant to many  features   of kinetic reconnection.

In the present study, we report results from a  series of simulations carried out by the authors within the scope of the preparations for the MMS mission \citep{sharma2005magnetospheric,curtis1999magnetospheric}. Appendix A reports the details of some of the key simulations used in the present study and of the method of simulation based on the iPic3D code~\citep{ipic3D}. The specific simulation reported is always indicated in the figure caption.

\begin{figure}
  \centering
  \includegraphics[width=\columnwidth]{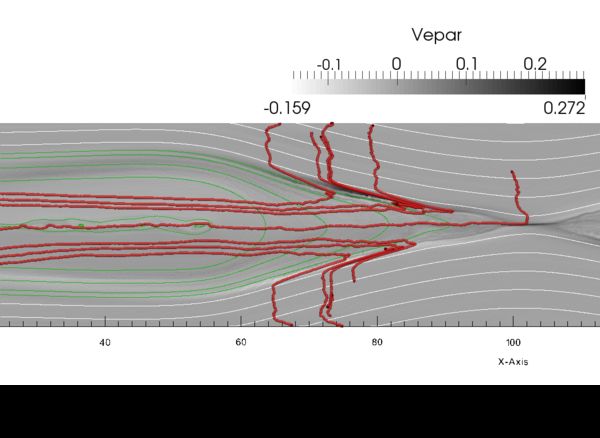}
  \caption{Run OpenBG1: State of the electron flow and of the magnetic field at a time $\omega_{ci}t=34.7$ (corresponding to cycle 35000). The electron flow lines are shown in red superimposed over the magnitude of the parallel electron speed (in grayscale). Additionally the magnetic field lines are shown in white for the family of lines not yet reconnected and in green for the field liens already undergone the process of reconnection. Only a portion of the domain is shown, including the full width in $y$ but a limited central range in $x$.  }
\label{eflow}
\end{figure}

\begin{figure}
  \centering
  \includegraphics[width=\columnwidth]{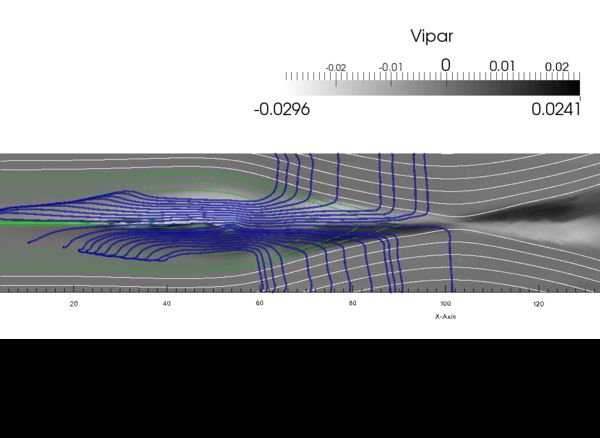}
  \caption{Run OpenBG1: State of the ion flow and of the magnetic field at a time $\omega_{ci}t=34.7$ (corresponding to cycle 35000). The ion flow lines are shown in blue superimposed over the magnitude of the parallel ion speed (in grayscale). Additionally the magnetic field lines are shown in white for the family of lines not yet reconnected and in green for the field liens already undergone the process of reconnection. Only a portion of the domain is shown, including the full width in $y$ but a limited central range in $x$.    }
\label{iflow}
\end{figure}

Figures \ref{eflow} and \ref{iflow} show the actual motion of the plasma species. Obviously, the description of the area around a reconnection site in terms of boxes is of limited use in understanding the electron and ion motion. The Petschek visualization is closer to what is observed, with the separatrices replacing the role of the slow shocks of the Petschek scheme. 

The flow lines shown in Fig. \ref{eflow}-\ref{iflow} are computed integrating the instantaneous average flow field in each cell. A Lagrangian fluid element would not necessarily follow such patterns because the flow field is subjected to fluctuations and changes. Similarly individual particles can deviate very significantly even from the Lagrangian motion of an element of plasma. Nevertheless, the instantaneous flow lines provide a picture of the flow field at a given time, being everywhere parallel to the velocity vector. 

The overall evolution of the magnetic field lines is also reported in Fig.  \ref{eflow}-\ref{iflow}. The initially horizontal field lines  bend vertically and move towards the central x-point where they are reconnected. The motion is caused by the $\bfE\times\bfB$ drift due to the out-of-plane reconnection electric field $E_z$. Subsequently, the newly reconnected field lines are advected downstream horizontally away form the x-point. 


The electrons and ions follow a pattern far different from the idealized image of the two nested boxes in Fig. \ref{boxes} representing the ion and electron diffusion regions. Both species are moved slowly, in a random-walk type of motion, towards the separatrix region where they become trapped into the faster conveyor belt of the reconnection process to be moved along the separatrix and inward across it. The electron motion is more sensitive to the parallel electric field pushing the electrons along the field lines towards the x-point. The ions, instead, are mostly moving across the separatrices. This difference is central to the physics of kinetic reconnection and will be addressed in the next section.

\begin{figure}
  \centering
  \includegraphics[width=\columnwidth]{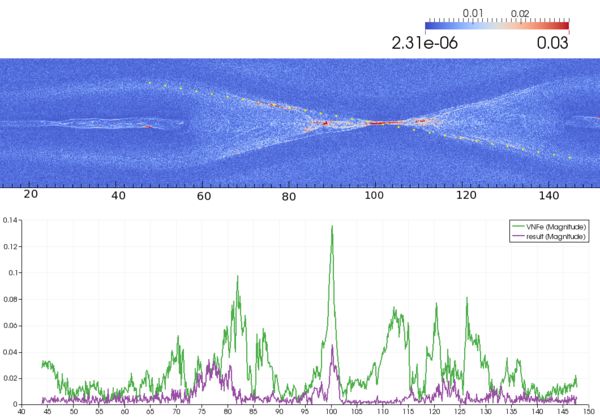}
  \caption{Run PerBG1: False color represenation of $V_{e\perp}- \bfE\times\bfB/B^2$. The top panel shows a blow up of the inner region and the bottom panel, a crossing along the line identified by the yllow dots in the top panel. The bottom panel shows both the total $V_{e}- \bfE\times\bfB/B^2$ (green) and the perpendicular part $V_{e\perp}- \bfE\times\bfB/B^2$ (violet). The separatrix has as strong a violation as the central electron diffusion region, albeit distributed over a more diffuse and less intense area.}
\label{unfrozen}
\end{figure}

The flow at the separatrices acquires a distinct non-frozen in component. The local magnetic field at the separatrices is significant, but the presence of strong gradients allow the electrons to violate the drift approximation, leading to a strong deviation from the local frozen-in condition. Figure~\ref{unfrozen} shows the relative drift between electrons and magneitc field lines, assuming the latter move at the $\bfE\times\bfB$ drift speed. The figure reports specifically the perpendicular component of the relative drift, because we have already observbed the parallel flow to be very strong there, but parallel flows do not consititute proper violation of the frozen in condition. Perpendicular drift do and the relative perpendicular drift at the seapratrices, even tens of $d_i$ away from teh x-point are still as strong as in the electron diffision region proper (the inner box in the nested boxes idealization of kinetic reconnection, see Fig.~\ref{boxes}).  

The violation of the frozen in condition extends downstream of the central electron diffusion region, forming a turbulent outflow that is subject to deflection \citep{Goldman:2011} caused by the guide field ($B_g=0.1B_0$, in the case reported) and by secondary tearing that forms secondary islands in the extended outflow from teh reconnection region~\citep{daughton2006fully}. 

Both species pass from the unreconnected region to the reconnected region mostly without even coming close to the central x-point. Whatever happens to them to make them change their distribution (see Sect.~\ref{sect-energy}) happens at the separatrices. The separatrices are in fact also the region where the electrons and ions acquire their energization.

The eventual outcome of reconnection is a front of plasma and piled up magnetic field exhaust from the reconnection region. This outflow motion is  observed often in satellite data and is referred to in that context as bursty bulk flow when reconnection events are transient or unsteady~\citep{angelopoulos1994statistical}. These flows inevitably meet the unperturbed plasma previously present and form a front downstream. These fronts also are often observed in satellite data and in the case of Earthward propagation contribute to restore the Earth field to a more dipolar configuration and for this reason are called dipolarization fronts \citep{sitnov2009dipolarization}.

\begin{figure}
  \centering
  \includegraphics[width=\columnwidth]{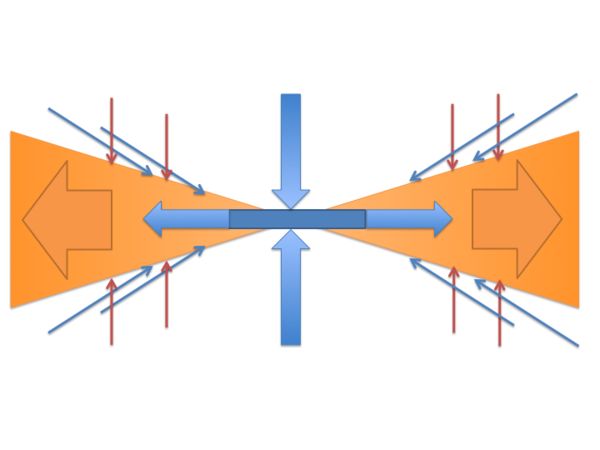}
  \caption{Cartoon of the actual electron and ion flow patterns in kinetic reconnection. Only a minority of electrons and ions pass near the x-point. Around this point, an electron diffusion region forms and an electron jets is emitted out. However, the vast majority of electrons and ions pass through the separatrices where their flow speed receives a large impulse over a narrow transition layer. The picture is more reminiscent of the Petschek scheme than of any Sweet-Parker box. The area within the separatrix (in orange) is filled with reconnected downstream plasma and the electrons an ions cross all separatrices.}
\label{boxseparatrix}
\end{figure}

Figure \ref{boxseparatrix} shows a cartoon of the actual flow pattern for the reconnection process and can replace the more idealised cartoon of Fig~\ref{boxes}. The difference from the two-boxes cartoon is profound, both electrons and ions primarily pass via the separatrix layer rather than the central electron diffusion region.

The four separatrices have the same flow  only in absence of guide field. The presence of guide field alters the flow pattern, but still in all cases the separatrices remain the gatekeepers. Already at the relatively modest 10\% level of Fig.~ \ref{eflow}-\ref{iflow}, the effect is quite visible on the electrons where a much stronger parallel flow is present on the upper left (and lower right, not shown) separatrix. The difference becomes stronger as the guide field is increased.

\begin{figure}
  \centering
  \includegraphics[width=\columnwidth]{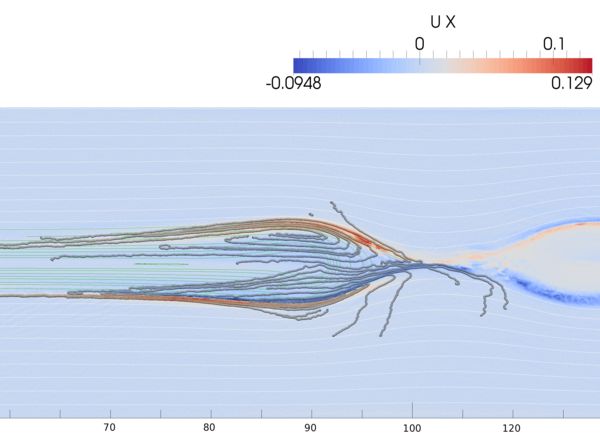}
  \caption{Run PerBG10: State of the electron flow and of the magnetic field at a time $\omega_{ci}t=19.7$ (corresponding to cycle 20000). The electron flow lines are shown in red superimposed over the horizontal component of the electron velocity. Additionally the magnetic field lines are shown in white for the family of lines not yet reconnected and in green for the field liens already undergone the process of reconnection. Only a portion of the domain is shown, including the full width in $y$ but a limited central range in $x$.  Note an earlier time is used to prevent the periodic boundary conditions from causing any effect.}
\label{eflowbg}
\end{figure}

\begin{figure}
  \centering
  \includegraphics[width=\columnwidth]{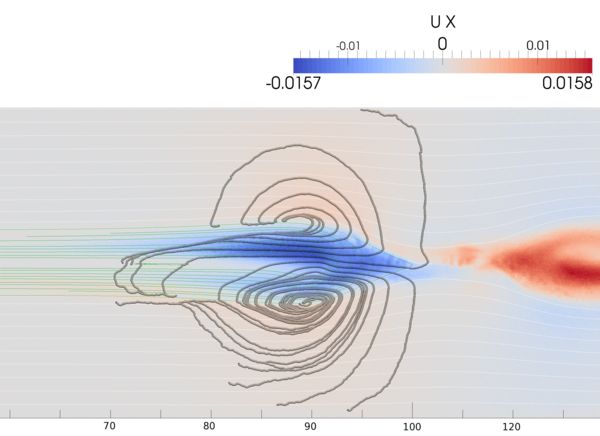}
  \caption{Run PerBG10: State of the ion flow and of the magnetic field at a time $\omega_{ci}t=19.7$ (corresponding to cycle 20000). The ion flow lines are shown in blue superimposed over the horizontal component of the ion velocity. Additionally the magnetic field lines are shown in white for the family of lines not yet reconnected and in green for the field liens already undergone the process of reconnection. Only a portion of the domain is shown, including the full width in $y$ but a limited central range in $x$.   Note an earlier time is used to prevent the periodic boundary conditions from causing any effect. }
\label{iflowbg}
\end{figure}

Figure \ref{eflowbg}-\ref{iflowbg} show, for $B_g=B_0$,  the flow lines again superimposed to the magnetic field lines. The flow is now of alternating sign. On two separatrices the parallel flow is still towards the x-point, but on the other two the flow becomes away from the x-point and strongly sheared  with two oppositely directed flow channels. As shown in Sect.~\ref{sect-instabilities}, these flow patterns are very susceptible to streaming and shear instabilities.  

The presence of a guide field also tends to give a more vortical motion to both plasma species, capturing them in  self-feeding reconnection conveyor belt loops~\citep{lapenta08,lapenta-lazarian}.

\section{Separatrices and signatures of the Hall physics}
\label{sect-hall}
The separatrices are strictly speaking lines in 2D and surfaces in 3D. However, within kinetic models, the separatrix proper become surrounded by a layer of repentine changes in a number of quantities, most prominent among them the electric and magnetic fields. The separatrices are highlighted by the in-plane perpendicular electric field and by the out-of-plane magnetic field. These two fields together are part of the characteristic signature of kinetic reconnection caused by the separation of scales between electrons and ions, represented mathematically by the Hall term in the generalized Ohm's law, and for this reason referred to as Hall fields.

 \begin{figure}
  \centering
  a)\\
  \includegraphics[width=.8\columnwidth]{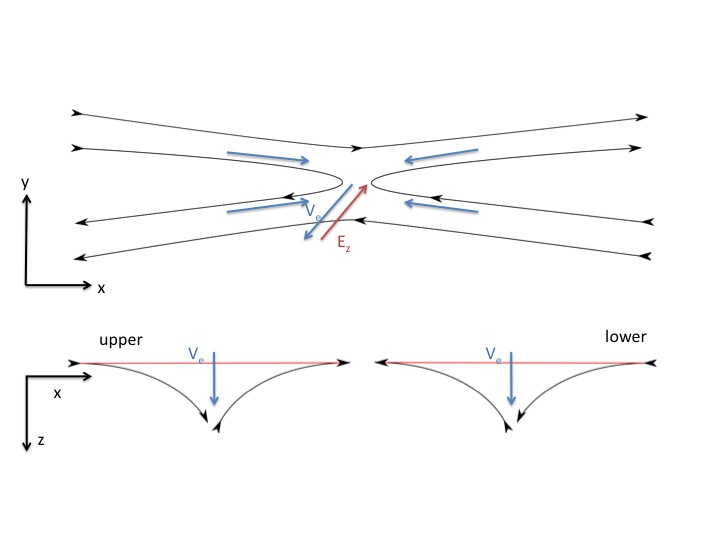}\\b)\\
    \includegraphics[width=.8\columnwidth]{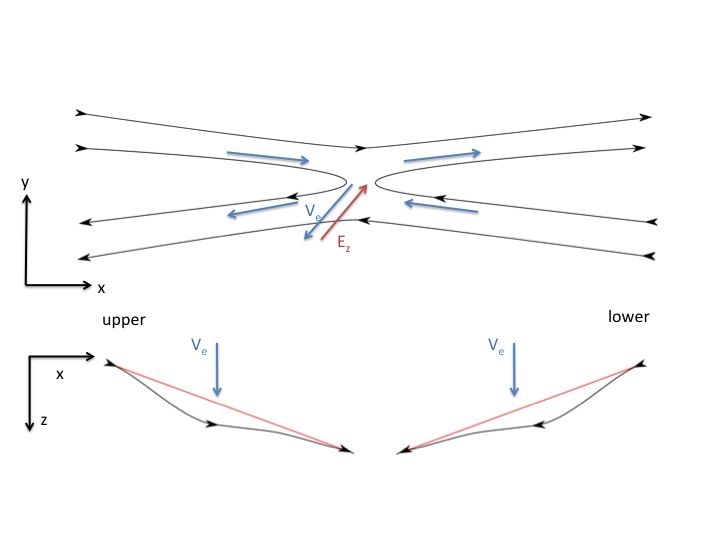}
  \caption{Hall physics in absence of guide field (a, top) and with guide field (b, bottom). Without guide field, the electron flow induced by the reconnection electric fiedl $E_z$ is directed alternatively in the opposite or in the same direction as that of the bent field lines, with a asymmetric up-down and left-right asymmetry. The out-of plane field is also doubly asymmetric, resulting in a parallel flow always directed towards the x-point. In presence of guide field, the field line bending perturbs the guide field without forming a quadrupolar pattern and the electron flow is always oppositely directed with the magnetic field, resulting in a alternating flow pattern moving towards the x-pint on two opposite separatrices and away from it on the other two.   }
\label{hall_physics}
\end{figure}

The concept of the Hall physics has been reviewed in previous studies (e.g. see the section by Shay and Drake in \citet{birn-priest}). A useful way of interpreting the dynamics is visualized  in Fig.~\ref{hall_physics}.

The reconnection process is associated with an out-of plane reconnection electric field $E_z$. The reconnection electric field, via the $\bfE\times\bfB$ drift, moves the magnetic field lines first towards the x-point and after reconnection, downstream away from it. Near the reconnection x-point, the electron beta becomes high and the electrons become capable of affecting the field line direction. This generates a mechanism where the electrons and the magnetic field attached to them acquire a strong out of plane velocity. The reconnection electric field $E_z$ tends to accelerate the electrons near the x-point~\citep{moses1993plasma,divin2010model}, along the $z$-direction, opposite to that of the reconnection electric field~\citep{huba2002three}. 

The field line bending is linked with a in-plane perpendicular electric field, the Hall electric field, that in proximity of the separatrices bends the field lines out of the page. The combined presence of a out-of-plane Hall magnetic field and in-plane Hall electric field is the main peculiarity of fast kinetic reconnection.  Field line bending and its associated electric and magnetic Hall fields propagates from the x-point along the magnetic field lines at superalfvenic speed~\citep{shay2011super,lapenta2013propagation}. The process has been interpreted as a kinetic \alfven wave (KAW) wing encompassing the seapratrices along all field lines affected by the electron acceleration along $z$~\citep{shay2011super,lapenta2013propagation}. 

This signature, in absence of an initial guide field assumes the peculiar and much publicised quadrupolar Hall magnetic field structure due to the bending indicated in Fig.~\ref{hall_physics}. However, when a initial guide field is added, the electrons do not need to bend the field lines in the same way as their z-directed motion becomes parallel, the more parallel the larger the guide field is. When the guide field is increased, the quadrupolar anti-symmetric signature is progressively replaced by a up-down symmetric signature (but still left-right antisymmetric) \citep{rogers2003signatures, gembeta}. 

Fig.~\ref{hall_physics} illustrates how the bending explains the symmetry of the out of plane magnetic field. With weak or no guide field, the left side acquires a out of plane field opposite to the right side and the upper side, opposite to the lower side, resulting in a quadrupolar field.  

The bending of the field lines explains also the parallel electron flow along the separatrices observed in the previous section. In absence of guide field the electron motion along $z$, normal to the initial field, acquires a parallel component once the field lines are bent out-of-plane. This allows a preferential route for the electrons along the bent lines, where they can travel at fast parallel speeds instead of the much slower $\bfE\times\bfB$ drift across field lines. The orientation of the electron flow and of the field lines results, as shown in Fig.~\ref{hall_physics}, in a flow always pointing towards the x-point. 

The situation changes radically in presence of guide fields~\citep{kleva1995fast,gembeta,lapenta2011bipolar}. When the guide field is present, the electron flow is always directed in the same orientation with respect to the field lines on both left-right sides of the x-point. However, it is opposite on the upper side when compared with the lower side. This results in the flow being directed towards the x-point on two separatrices and away from it on the other two. 

The fast parallel electron motion results also in a tendency to create a charge imbalance that produces a density perturbation on the separatrices, and being the electrons the main current carriers, a violation of neutrality results in a narrow band of net charge and strong electric fields at the separatrices, constituting the source of the Hall electric  field, primarily electrostatic in nature, that is needed as part of the kinetic alfven wave wing forming the Hall signature along the separatrices \citep{wygant2005cluster, zhi2008generation}.

\begin{figure}
  \centering
  a)\\
  \includegraphics[width=\columnwidth]{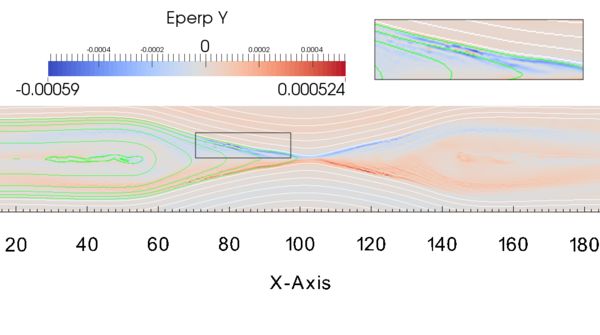}\\b)\\
    \includegraphics[width=\columnwidth]{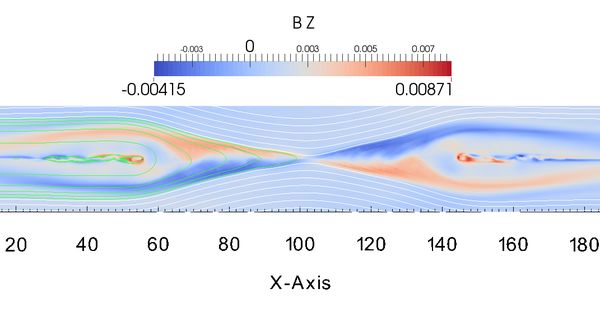}
  \caption{Run OpenBG1: Hall signatures  at time $\omega_{ci}t=34.7$ (cycle 35000) for the run described in the text. Top (a): $y$ component of the perpendicular electric field. A subset area near one separatrix is blown up on the top right corner, to highlight features not easily visible on the full figure. Bottom (b): Out of plane component of the  magnetic field.  }
\label{hall_signatures}
\end{figure}

\begin{figure}
  \centering
  a)\\
  \includegraphics[width=\columnwidth]{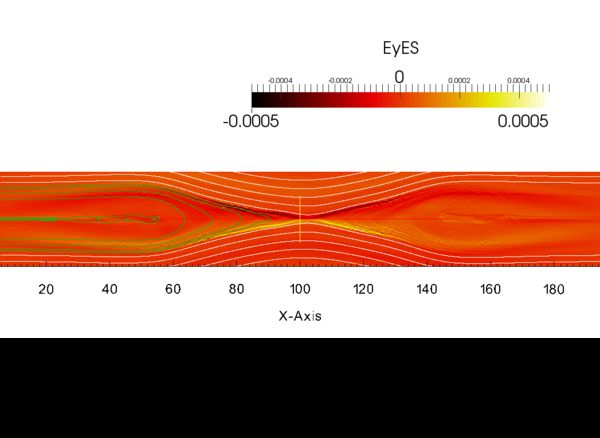}\\ b)\\
    \includegraphics[width=\columnwidth]{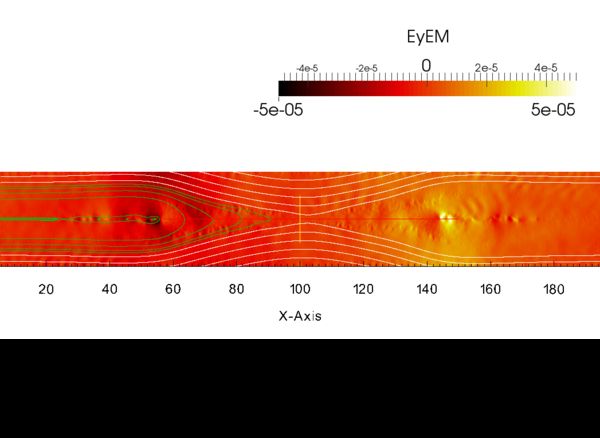}
  \caption{Run OpenBG1: Electrostatic (a,top) and electromagnetic (b, bottom) contributions to the electric field component $E_y$ for the same case in Fig.~\ref{hall_signatures}.  }
\label{hallE}
\end{figure}

Figure \ref{hall_signatures} shows the Hall signatures for  the electric and magnetic fields for low guide fields ($B_g=0.1B_0$). The in-plane perpendicular electric field is caused by a charge imbalance across the separatrix and is electrostatic in nature \citep{wygant2005cluster, zhi2008generation, PLA:400587}. The out of plane magnetic field is caused  by an electron current system flowing along the separatrix~\citep{uzdensky2006physical}.  Figure \ref{hallE} reports the electrostatic and electromagnetic contributions to the vertical in-plane component of the electric field. The electrostatic contribution is obtained from the scalar potential defined by the Poisson's equation, the electromagnetic component can be obtained from the vector potential and is simply the reminder of the total electric field minus the electrostatic component. As can be observed the electromagnetic component is an order of magnitude smaller and does not show any specific feature at the separatrix, confirming the electrostatic nature of the Hall electric field.

\begin{figure}
  \centering
  a)\\
  \includegraphics[width=\columnwidth]{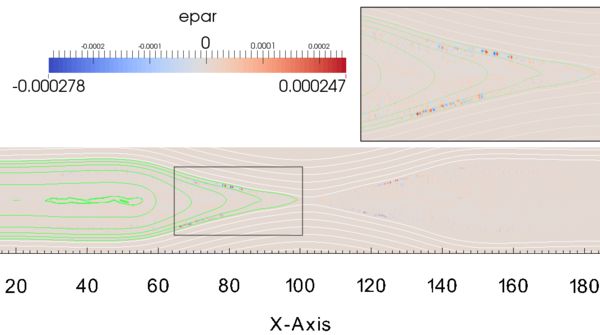}\\ b)\\
    \includegraphics[width=\columnwidth]{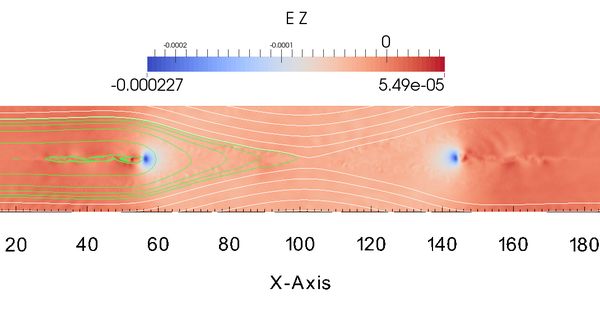}
  \caption{Run OpenBG1: Other key reconnection signatures at time $\omega_{ci}$ for the same state shown in Fig.~\ref{hall_signatures}. Top (a): Parallel electric field (a scalar, $\bfE\cdot\bfB/B$). A subset area near one separatrix is bown up on the top right corner, to highlight features not easily visible on the full figure. Bottom (b): Reconnection electric field, out of plane.      }
\label{other_signatures}
\end{figure}

Figure \ref{other_signatures} shows the parallel electric field characteristic of the separatrix and the reconnection electric field, controlling the rate of reconnection. At this late time, the reconnection electric field is dominant in the so-called dipolarization fronts forming a flux pileup region downstream of  reconnection, where the plasma outflowing from reconnection encounters the fresh original still unperturbed plasma~\citep{sitnov2009dipolarization}. The parallel electric field shows clearly the presence of bipolar structures associated with electron holes caused by streaming instabilities, the topic of the next Section.

\
\section{Instabilities of reconnection separatrix layers}
\label{sect-instabilities}

Fast bulk flows and localized electron beams, formed by reconnection, are inherently unstable to current-driven and beam-driven modes. Electron holes (EH), whistler waves, lower-hybrid oscillations, double layers, small-scale plasma vortices, and other dynamic phenomena are observed reconnection events in satellite data and simulations (see review by \citep{Fujimoto2011}, also references therein). Magnetic reconnection in space plasma easily generates a variety of multiscale non-stationary features, even if the process retains the common global X-line, separatrices, inflow and outflow regions.

In the past decades, significant efforts were put into revealing the nature of the collisionless plasma dissipation, which is required for violation of frozen-in constraint near the X-line. Kinetic theory of microinstabilities suggests that wave-particle interactions produce turbulent drag, creating ‘anomalous’ resistivity (\citep{Galeev84anomalous, biskamp}) is orders of magnitude larger than Spitzer resistivity due to binary collisions \citep{Spitzer}.

Three-dimensional PIC simulations are essential for understanding the collisionless dissipation. A number of studies suggest that various microinstabilities are excited inside diffusion regions, among them: (\textit{e}-\textit{e}) streaming instability \citep{Goldman2008}, (\textit{e}-\textit{i}) Buneman instability \citep{Drake2003}, (\textit{e}-\textit{e}-\textit{i}) Lower-Hybrid instability, excited by a propagation of a suprathermal electron beam \citep{Che2009, Che2010}, unstable electron cyclotron mode \citep{Che2011}, excited by strong current density gradient \citep{Drake1994} inside diffusion region. The anomalous drag is found by averaging the electron motion equation over the current direction and estimating the correlations between fluctuations in the density, electric and magnetic fields and the velocity.

The problem of collisionless dissipation and anomalous resistivity remains one of the unsolved mysteries of magnetic reconnection. However, as has been shown in previous Sections, fast flows and beams are not confined to the X-line vicinity, and exist many $d_i$’s away as e.g. a part of large-scale Hall region current system, or as travelling reconnection jet fronts. Therefore, it is important to understand the wave-particle interactions, which appear at large distances from the X-line.

Early 2D PIC simulations of antiparallel reconnection \citep{Hoshino2001} reported bi-Maxwellian distributions at reconnection separatrices (a weak electron jet superimposed on thermal electron population). It was found that the wave activity at the separatrices is enhanced in a broad range of frequencies between the lower hybrid frequency and electron plasma frequency. Similar electron distribution functions at separatrices were reported in Ref. \citep{Pritchett2004}. The study found waves with the frequency close to the local ion plasma frequency, interpreted as evidence of either Buneman, or electron streaming modes. 

Traditionally, Hall physics is considered to be important in the vicinity of the ion diffusion region, where ion and electron motion decouples \citep{birnGEM}. However, as shown in the previous sections, Hall signatures at separatrices spread out up to $\sim 100$ $d_i$’s. away from neutral line. 2D PIC simulations performed in larger boxes \citep{Pritchett2005, Cattell2005, Drake2005} allowed to simulate remote separatrices. The parallel electric field creates a beam of electrons susceptible to the Buneman instability. With the typical mass ratio of 64 or 100 used in many PIC simulations, the disparity of ion and electron scales is not resolved well, hence the properties of sub-ion instabilities and structures are not modeled well. Numerical simulations with the realistic mass ratios \citep{Lapenta2010, lapenta2011bipolar, Divin2012} allowed to establish the inner layered structure of the separatrices and identify the Buneman instability and electron Kelvin-Helmholtz mode as the basic drivers of separatrix wave activity.

\begin{figure*}
\includegraphics[width=\textwidth]{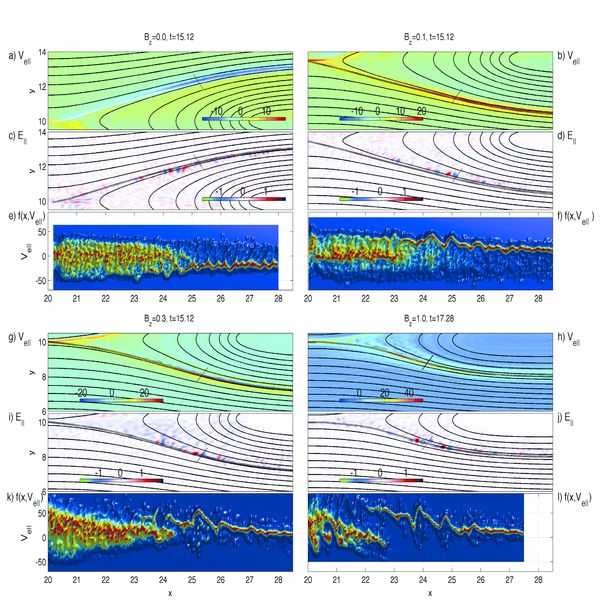}
\caption{Overview of the 2D reconnection simulations, one quadrant is presented from each simulation. The parallel electron velocity $v_{e||}$ (panels a, b, g, h). The parallel electric field $E_{||}$ (panels c, d, i, j). Electron phase space density $f(x,v_{e||})$ along the separatrix (sampling line is shown with thin black curve). The panels are grouped in 3 plots to represent, respectively (from left to right and from up to down): $B_z=0$, $B_z=0.1$, $B_z=0.3$, $B_z=1.0$ runs. }
\label{fig:Fig1}

\end{figure*}

\newpage

\begin{figure}
\centering
\includegraphics[width=\textwidth]{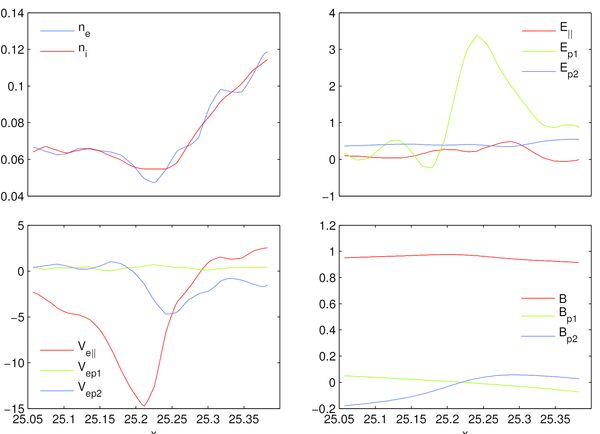}
\caption{Plasma properties along the cut marked in Fig. \ref{fig:Fig1}a with respect to $\textbf{x}$ axis ($B_z=0$ run). The profile is perpendicular to the magnetic field lines (the plots have different scales of $x$ and $y$ grid). The separatrix and the profile intersect at $x \sim 25.2$.  Field-aligned coordinate system is $\vec{e}_{||} = \vec{B}/|\vec{B}|$, the perpendicular vectors are $\vec{e}_{p1}=(\vec{e}_{z} \times \vec{e}_{||})/|\vec{e}_{z} \times \vec{e}_{||}| $, $\vec{e}_{p2}=\vec{e}_{p1} \times \vec{e}_{||}$. }
\label{fig:Fig2}
\end{figure}

\begin{figure*}
\includegraphics[width=\textwidth]{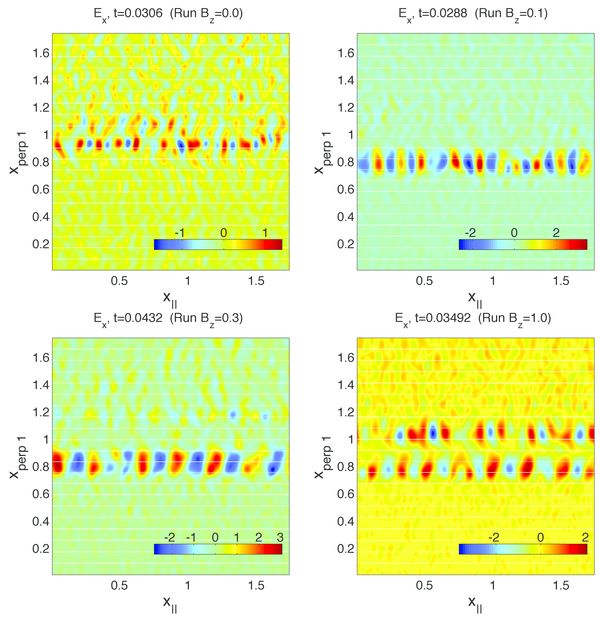}
\caption{Distribution of the parallel electric field along the separatrix flow channel in reduced-size simulations. Simulations are performed within the plane ($x_{||}-x_{p1}$), in which the growth of the Buneman instability is observed.}
\label{fig:Fig3}
\end{figure*}

The main goal of this section is the investigation of the instabilities developing at the separatrices of collisionless guide-field magnetic reconnection with realistic mass ratio simulations. In addition to conventional 2D runs \citep{Lapenta2010, lapenta2011bipolar}, we performed a set of small-scale high-resolution runs, in which the separatrix is approximated by a thin electron-scale current layer superimposed with a fast electron beam. This approach was used previously in Ref. \citep{Divin2012} to investigate separatrices of $B_z=1$ guide field case. These simulations found the Buneman instability developing parallel to the magnetic field, and electron Kelvin-Helmholtz (EKH) mode, which is excited by a weak shear of the electron velocity inside the separatrix density cavity. To complement the study, we performed a set of simulations with the guide fields $B_z=0.1$, $B_z=0.3$, $B_z=1$, including small-scale high-resolution runs of the dynamics of the separatrices. The rest of the section is organized as follows. Next we discuss the guide-field simulations of magnetic reconnection and the properties of plasma at separatrices. After that we present the results of the small-scale high-resolution 2D runs. After that we provide discussion and summary of our simulations.

The simulations are initialized with the Harris current sheet \citep{Harris1962}, which have the temperature of ions $T_i$ to electrons $T_e$ ratio of $T_i/T_e=5$, which is typical for magnetotail studies \citep{Lapenta2010}. The computational box is $1024$ $\times$ $512$ grid point. The physical size of the computational domain $L_x \times L_y = 40 d_i \times 20 d_i$. Periodic boundary conditions are set in the $\textbf{x}$ direction. PEC (Perfect Electric Conductor) boundaries are used at $y= 0$ and $y=L_y$. Alfv\'en dimensionless units are used in this section to visualize the results.

The initial configuration (Harris equilibrium) evolves into an asymmetric reconnection configuration in the presence of guide field. The vicinity of the reconnection region is antisymmetric, that is two separatrices (out of 4) are characterized by increased plasma density (denoted as “quiet”), and the other two separatrices (denoted as “active” \citep{Divin2012}) host thin density cavities, which contain intense electron jet.

Results of all the four simulation runs are presented in Fig. \ref{fig:Fig1}. We show only one quadrant of each simulation in order to have a compact comparative picture. No guide-field run ($B_z=0$, panels \ref{fig:Fig1}a, \ref{fig:Fig1}c, \ref{fig:Fig1}e) has symmetric reconnection pattern, therefore we can choose any separatrix for the detailed study. Guide-field runs have sharply different active and quiet separatrices, therefore we have chosen right lower quadrant, which contains the active one. 

Fig. \ref{fig:Fig1} shows the parallel electron velocity $v_{e||}$ (Fig. \ref{fig:Fig1}a, \ref{fig:Fig1}b, \ref{fig:Fig1}g, \ref{fig:Fig1}h), the parallel electric field $E_{||}$ (Fig. \ref{fig:Fig1}c, \ref{fig:Fig1}d, \ref{fig:Fig1}i, \ref{fig:Fig1}j), and the parallel electron velocity phase space $f(x,v_{e||})$ plot along the field line passing through the X-line, which is located at $x \sim 20$, $y \sim 10$. A curve on each 2D plot shows a portion of the computational domain, from where the electron distribution functions for the phase space plots (Fig. \ref{fig:Fig1}e, \ref{fig:Fig1}f, \ref{fig:Fig1}k, \ref{fig:Fig1}l) were sampled. Thin lines at $x \sim 25$ mark the profiles used in small-scale simulations later. Black curves are lines of the in-plane $(B_x,B_y)$ magnetic field.

Reconnection separatrices host electron flow layer, which spans many ion inertial lengths and has a large parallel velocity $V_{||}$.  Peak parallel electron velocity is $~10$ in the $B_z=0$ case and increases with $B_z$, reaching $\sim 45V_A$, which is comparable to the electron Alfv\'en velocity $(m_i/m_e)^{1/2}\sim 42$. Electrons are magnetized at separatrices, because the magnetic field is almost equal to the ambient value $|\vec{B}| \sim 1$ far from the reconnection region.

The thin electron flow layer creates a large shear in the parallel electron velocity across the separatrix. As discussed previously in Ref. \citep{Divin2012}, electron flow layer is unstable to the EKH mode for the $B_z=1$ case. Later in this section we show that the simulations with $B_z=0$, $B_z=0.1$, $B_z=0.3$ also create unstable electron flows at the separatrices.

The plots of the parallel $\textbf{E}$ field show the presence of the strong electrostatic fluctuations reaching $E_{||} \sim E_A$, which are commonly interpreted as electron holes \citep{Pritchett2005, Cattell2005, Drake2005}. The amplitude of the $E_{||}$ fluctuations are similar among all 4 runs, unlike the electron beam velocities. Notably, the EH size is slightly larger than the electron flow layer thickness, meaning that the EH electric field leaks into the surrounding plasma. This is seen in Fig. \ref{fig:Fig1}h, \ref{fig:Fig1}j ($B_z=1$ run), where the flow layer is particularly thin. In order to clarify further the separatrix layer properties and stability, we investigate the phase space densities (PSD).

Sampling boxes used for the PSD plots (Fig. \ref{fig:Fig1}e, \ref{fig:Fig1}f, \ref{fig:Fig1}k, \ref{fig:Fig1}l) are small ($0.04$ $d_i$ $\times$ $0.04$ $d_i$) in order to avoid the overlapping with the bulk slow plasma outside the electron flow layer. The PSD plots show the presence of the two rather different electron populations. Hot population exists in the area $20<x<23$, where the X-line acceleration is important. At $x > 23$, a strong electron jet is visible in all four cases, with the velocity peaking at $\sim 60 V_A$. Fluctuations of the $E_{||}$ coincide with PSD perturbations typical for the electron holes. The jet contains a small thermal electron population ($B_z=0$, $B_z=0.1$ runs), but nearly all background plasma is evacuated from the flow channel in the other runs ($B_z=0.3$, $B_z=1.0$). Electron holes are most prominent in the jet area $x > 23$, and much fainter close to the X-line ($20<x<23$)

Next we investigate the electron flow layer stability using reduced size simulations \citep{Goldman2008, Divin2012}. We are interested in the dynamics of quasi-1D separatrix layer only, and we sample distribution functions and plasma parameters from that region only and initialize small-scale high-resolution simulations of the size $1.75$ $d_i$ $\times$ $1.75$ $d_i$. This approach allows to use rotated coordinate system, therefore it is also possible to study instabilities with a wavevector in the \textbf{z} direction. The profile is shown in Fig. \ref{fig:Fig1} by a thin black line. 

The following new field-aligned frame of reference is introduced: the unit vector parallel to the $\vec{B}$ field is $\vec{e}_{||} = \vec{B}/|\vec{B}|$, the perpendicular vectors are $\vec{e}_{p1}=(\vec{e}_{z} \times \vec{e}_{||})/|\vec{e}_{z} \times \vec{e}_{||}| $, $\vec{e}_{p2}=\vec{e}_{p1} \times \vec{e}_{||}$. The vector $\vec{e}_{p1}$ is parallel to the profile line showin in Fig. \ref{fig:Fig1} and points across the separatrix.

Plasma parameters in the field-aligned frame of reference are shown in Fig. \ref{fig:Fig2} for $B_z=0$ run with respect to $\textbf{x}$ axis. The actual profile length is $0.9$ $d_i$, and it is centered at the density cavity. Unlike $B_z=1$ case (see Ref. \citep{Divin2012}), the density drop is not deep (the lowest density is $n_e \sim 0.05$. Visible difference of $n_e$, $n_i$ corresponds to a strong perpendicular electric field, which represents the standing whistler wave during fast collisionless reconnection. A small $E_{||}$ is present inside the cavity and accelerates the beam electrons along the separatrix up to $\sim 15$ $V_A$. Notably, a strong electron flow reversal of $v_{ep2}$ is responsible for the EKH mode excitation (see below).

The profile parameters presented in Fig. \ref{fig:Fig2} are used to initialize the 2D PIC simulations in the domain $1.75$ $d_i$ $\times$ $1.75$ $d_i$. The areas $0<x_{p1}<0.45$ and $1.35<x_{p1}<1.75$ contain the uniform plasma which corresponds to the left and right sides of the profile in order to allow for a buffer layer between the electron flow jet and the boundary. The computational domain has Perfect Electric Conductor (PEC) boundaries at the $x_{p1}=0$,  $x_{p1}=1.75$, whereas the other direction is periodic. The simulations are performed in the plane $x_{||}-x_{p1}$, which resolve the Buneman instability, and in the plane $x_{p2}-x_{p1}$, which sees the development of EKH instability, in agreement with the simulations performed in Ref. \citep{Divin2012}.

\begin{figure}
\centering
\includegraphics[width=0.75\textwidth]{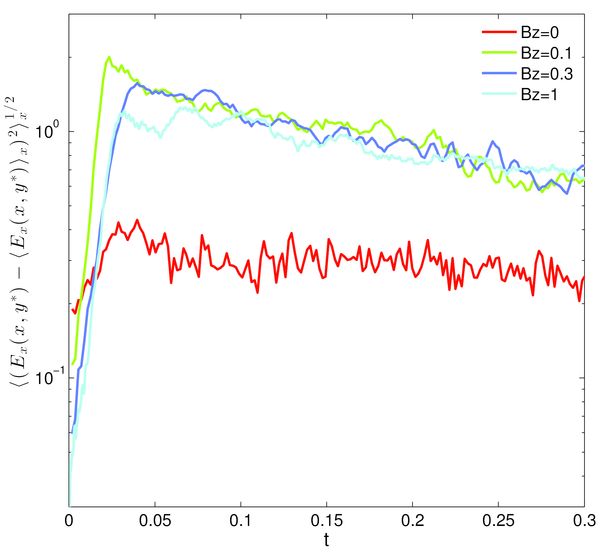}
\caption{Evolution of  $(\langle E_{||}(x_{||},x_{perp1}^{*},t)  - \langle E_{||}(x_{||},x_{perp1}^{*},t) \rangle_{x||}    \rangle_{x||})_{x||}^{1/2}$ for all 4 runs. The value of $x_{perp1}^{*}$ is taken at a line, where the most intense $E_{||}$ pulsations are observed, see Fig. \ref{fig:Fig3}. $x_{perp1}^{*}$ satisfies $0.8<x_{perp1}^{*}<1$ for all the four runs.}
\label{fig:Fig4}
\end{figure}

\begin{figure}[h]
\centering
\includegraphics[width=0.75\textwidth]{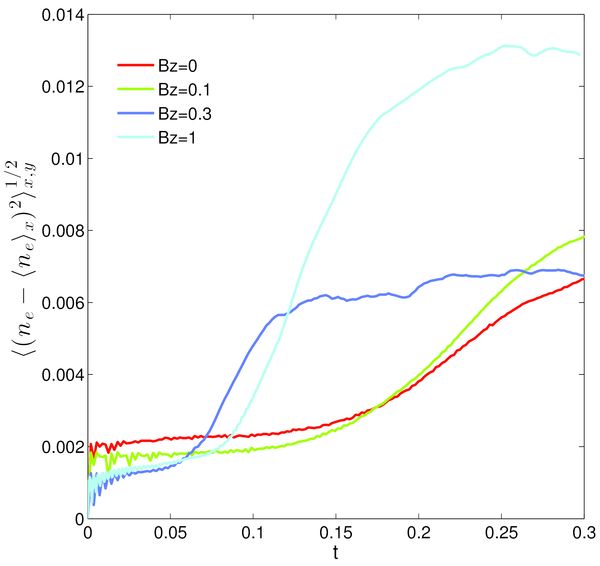}
\caption{Evolution of  $(\langle (n_{e}  - \langle n_{e} \rangle_{xp2})^2    \rangle_{xp2, xp1})^{1/2}$ for all 4 runs.}
\label{fig:Fig5}
\end{figure}

\begin{figure*}
\includegraphics[width=\textwidth]{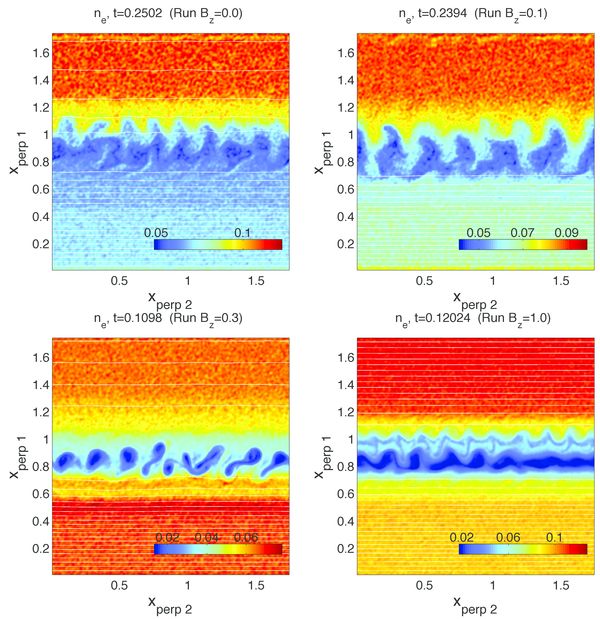}
\caption{Distribution of the electron density along the separatrix flow channel in reduced-size simulations. Simulations are performed within rotated ($x_{p2}-x_{p1}$) plane, in which the growth of the electron Kelvin-Helmholtz mode is observed.}
\label{fig:Fig6}
\end{figure*}

The electron holes found in magnetic reconnection simulations in Fig. \ref{fig:Fig1} are readily observed in small-scale runs. Fig. \ref{fig:Fig3} displays $E_{||}$ component taken from the 2D small-scale simulations performed within the plane ($x_{||}-x_{p1}$) for the four guide field cases. A specific time corresponds to a peak of the norm $(\langle E_{||}(x_{||},x_{p1}^{*},t) - \langle E_{||}(x_{||},x_{p1}^{*},t) \rangle_{x||} \rangle_{x||})_{x||}^{1/2}$, where $\langle \rangle_{x_{||}}$ denotes averaging along $x_{||}$ direction, computed along the middle of the jet. As discussed previously in Ref. \citep{Divin2012}, multiple electron jets can be found near the separatrices (as seen in Fig. \ref{fig:Fig3}, $B_z=1.0$, $x_{p1}\sim 1.05$), where a secondary jet is well visible.  The larger the guide field, the faster is the electron flow jet (see $v_{e||}$ plots in Fig. \ref{fig:Fig1}). Respectively, stronger electrostatic waves are expected for $B_z>0$. Fig. \ref{fig:Fig3} indicates that the peak $E_{||}$ indeed grows with the guide field. 

Time evolution of the $E_{||}$ wave intensity is displayed next in Fig. \ref{fig:Fig4}. Instability growth, saturation and nonlinear evolution can be best visualized by plotting the norm $(\langle E_{||}(x_{||},x_{p1}^{*},t) - \langle E_{||}(x_{||},x_{p1}^{*},t) \rangle_{x||} \rangle_{x||})_{x||}^{1/2}$ evolution with time. Notably, $B_z=0$ run (Fig. \ref{fig:Fig4}, red line) has significantly weaker wave amplitude and growth rate ($\gamma_{Bz0} \sim 22$) compared to $B_z>0$ runs ($90<\gamma<120$). This may indicate that even a small guide field is sufficient to noticeably enhance the separatrix electrostatic wave activity. However, in actual large-scale reconnection a dynamical equilibrium exists between the EH generation, saturation and dissipation \citep{Goldman2014}. Therefore, it is uneasy to find a “growing” EH in the large-scale simulations, and linear dispersion relations can only provide very rough estimate of the actual wave properties.

Next we discuss the results of 2D PIC simulations performed in the plane $(x_{p2}-x_{p1})$. As we discovered in our previous study \citep{Divin2012}, the electron Kelvin-Helmholtz mode is excited for the case $B_z=1.0$. A weak unstable shear flow exists in the $x_{p2}$ direction, leading to electron vortices in the plane $(x_{p2}-x_{p1})$, which fill the density cavity with inflow and outflow plasma. Fig. \ref{fig:Fig5} displays electron density from all four small-scale runs. Specific times correspond to a stage, in which the value of the norm $(\langle (n_{e} - \langle n_{e} \rangle_{xp2})^2 \rangle_{xp2, xp1})^{1/2}$ is $\sim 0.0055$ in all four runs (see Fig. Fig. \ref{fig:Fig6}), indicating a comparable level of fluctuations. Notably, the density cavity is nearly absent in the case $B_z=0$, hence EKH mixes plasma from inflow and outflow sides. The same applies to the case of $B_z=0.1$. However, for $B_z=0.3$ and $B_z=1.0$ the density inside the cavity is $\sim 0.02$, and the EKH produces significant mixing of the cavity electron population, and inflow/outflow populations. It indicates that the EKH instability is especially important in the large guide field case and can divide the separatrix surface into a chain of isolated beams.

\section{3D study of separatrix instabilities}\label{sec:3dlociofinstability}

\begin{figure}
  \centerline{\includegraphics[width=1.0 \columnwidth]{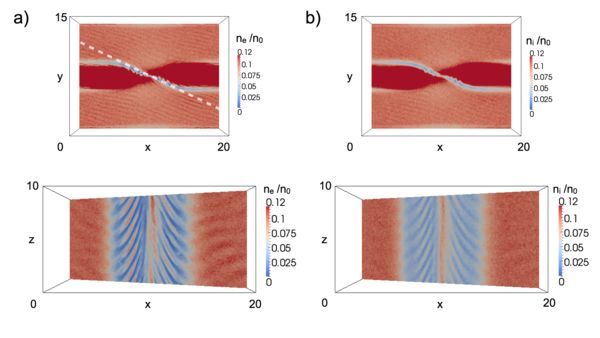}}
  \caption{Contour plots of the electron (panel a) and ion (panel b) density on the reconnection plane and on a plane along the separatrices. The plot shows the density cavity layers (blue regions), and the presence of density ripples, reminiscent of Kelvin-Helmholtz vortices, along the cavities. The local magnetic field in the cavity is approximately $45^\circ$ out of the presented plane. The contour plot on the plane along the separatrices shows the presence of {\em low density ribs}.}
\label{fig:ka}
\end{figure}

One of the distinctive characteristics of separatrices in collisionless magnetic reconnection is the development of localized low density regions along them~\cite{Lu:2010,Zhou:2011}. These depletion areas are called {\em cavities}. The cavities appear as thin electron current layers in two dimensional Particle-in-Cell simulations and as thin surfaces in three-dimensional simulation~\cite{markidis2012three}. Figure \ref{fig:ka} shows the contour plot of electron (panel a) and ion (panel b) on two different planes. The top panels shows the contour plot on the reconnection plane while the bottom panels show a contour plot on a plane along the separatrices (dashed line in panel a). The density cavities appear in blue as depleted areas along the separatrices.

One of the findings of studies about cavities in three-dimensions is that regions with further lower density are embedded in the cavities~\cite{markidis2012three,wang2013observation}.  As clear in the bottom panels of Figure~\ref{fig:ka}, these structures are similar in shape to ribs, and for this reason these regions are called {\em low density ribs}. Separate low density ribs show similar features. They have equal size, they form along the magnetic field lines and are supported by intense perpendicular electric fields. A parallel electron current moves along the cavities and contributes to the creation of the low density ribs by decreasing the density in small channels. The cause of the three-dimensional low density ribs is not clear. It has been proposed that lower-hybrid waves can be caused by electron beams~\cite{drake science,McMillan:2007,Che2009,Che2010}, or by density gradients, as in the lower-hybrid drift instability~\cite{Scholer:2003}. The density ripples in the cavities have been previously detected in two dimensional Particle-in-Cell simulations, and spectral analysis suggests that the instability is related to the electron MHD Kelvin-Helmholtz mode \cite{Divin2012}. An additional possibility is that observed three-dimensional structures of higher density regions in proximity of the low density ribs are Alfv\'en vortex filaments~\cite{Alexandrova:2006,Xu:2010}, caused by drift kinetic Alfv\'en waves. The presence of Alfv\'en vortices have been detected in cusp region~\cite{Sundkvist;2005} and close to magnetic reconnection X lines~\cite{Chaston:2005, Chaston:2009}.

It is clear from the top panels of Figure~\ref{fig:ka} that cavities are perturbed by a wave structure \cite{Divin2012}. Small spatial scale low density ripples develop along the cavities. These structures are similar in shape to the Kelvin-Helmholtz vortices \cite{Divin2012}. 

\begin{figure}
  \centerline{\includegraphics[width=1.0 \columnwidth]{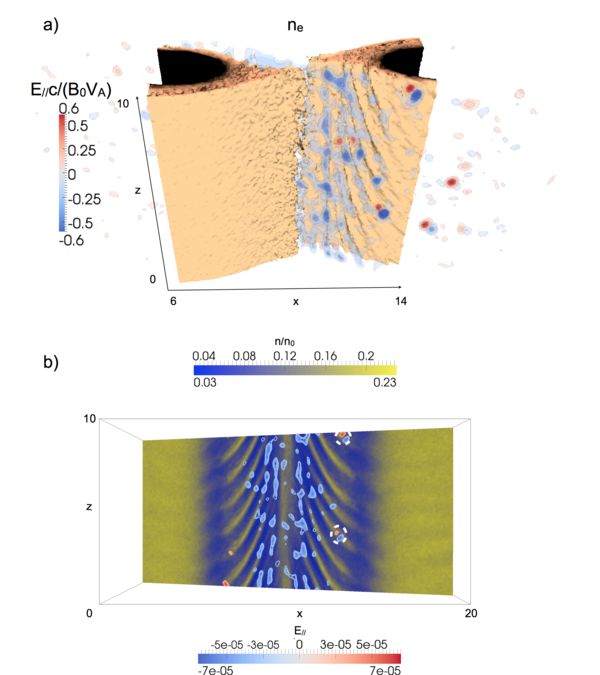}}
  \caption{ Electron density isosurface plot orange and grey colors with parallel electric field isosurface plot for $E_{//} = \pm 0.3 \ B_0V_A/c$ in red and blue colors during magnetic reconnection in panel a. In panel b, an isosurface plot of $E_{//} = \pm 0.3 \ B_0V_A/c $ is superimposed to a contourplot of the total density along a separatrix. The bipolar parallel electric field  structures develop along the low density ribs and magnetic field lines direction between the cavities and the outflow regions.}
\label{fig:kb}
\end{figure}

Bipolar parallel electric field ($E_{//} =  \mathbf{E} \cdot \mathbf{B}/|\mathbf{B}|$) signatures are present along the cavities. Figure\ref{fig:kb} shows an isosurface plot of $E_{//} = \pm 0.3 \ B_0V_A/c $ with density isosurfaces that represent to outflow jets (panel a) and a contourplot of the total density along a separatrix. There is correspondence between the electric field bipolar structures and the {\em low density ribs}. The existence of the low density ribs within the cavities and of bipolar electric field structures can possibly used as flag to detect the reconnection sites.

The instabilities observed in 3D simulations are sensitive to the ambient guide field. As the guide field is increased, the flow pattern around the separatrices changes, as observed in Sect.~\ref{sect-flow}. These flows are also a main driver for the instabiltiies at the separatrices, creating a direct link between the ambient guide field and the presence of different types of instabilities. 

\begin{figure}
\centering
\includegraphics[width=\linewidth]{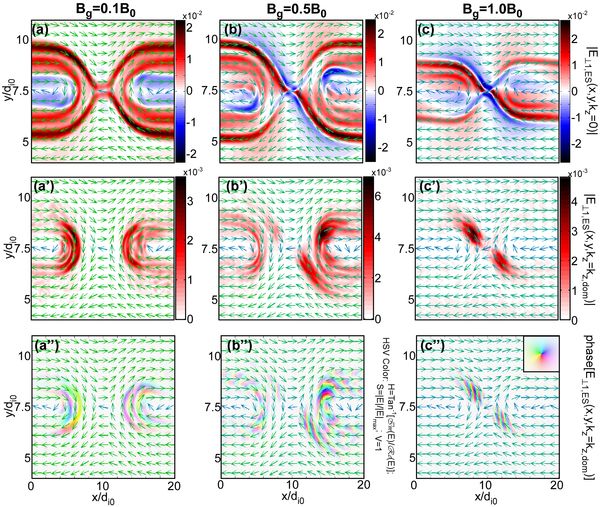}
\caption{Selected Fourier modes of $E_{ES\perp1}(x,y,k_z)$ from 3D simulations with guide fields $B_g/B_0=$ 0.1 (a)--(a''), 0.5 (b)--(b''), and 1.0 (c)--(c'').  The top row (a)--(c) is the $k_z=0$ mode (equal to the $z$ average). The second row (a')--(c') and third row (a'')--(c'') are, respectively the magnitude and complex phase of the $k_z\ne 0$ mode with the largest maximum  magnitude for each value of $B_g$, as specified in the space between the second and third row. The HSV color representation of the amplitude weighted complex phase is keyed to the inset in the upper-right corner of (c''). Results from runs: 3DPeriodicBG1, 3DPeriodicBG5, 3DPeriodicBG10.}
\label{fig:Bg-comp}
\end{figure}

As the guide field is increased the paralell flow at the separatrices becomes more intesnse and more susceptible to streaming instabilities~\citep{Lapenta2010}. However, guide fields tend to provide a path of communication along the out of plane direction $z$, suppresing some of the instabilities that develop in htat direction. A transitionis then observed as a function of the guide field. At low guide fields, the dominant presence is that of instabilties in the lower hybrid range at density gradients, observed in the flanks of the current sheet~\citep{lapentabrackbill02} and at the dipolarization front~\citep{vapirev2013formation}. At higher guide fields, the separatrix instabilities become more dominant, as shown in Fig.~\ref{fig:Bg-comp}.

Figure \ref{fig:Bg-comp} provides a comparison of the perturbations in $z$ of $E_{ES\perp1}$ for three different values of the guide field $B_g/B_0=$0.1, 0.5, and 1.0. Here, the $\perp\!\!1$ direction is parallel to $\mathbf{B}\times\hat z$, and is therefore the direction $90^\circ$ clockwise from $\mathbf{B}$ projected into the $x$--$y$  plane. The electrostatic part of $\mathbf{E}$ is defined as $\mathbf{E}_{ES}=\nabla[\nabla^{-2}(\nabla\cdot \mathbf{E})]$.     

The simulation data is taken from a series of runs over a range of guide-field strengths in a box of dimension $(L_x,L_y,L_z)=(20,15,10)d_{i0}$, all at time $\Omega_it=14.6$.  The 3D array $E_{ES\perp1}(x,y,z)$ is Fourier transformed along the $z$ direction only, to generate $E_{ES\perp1}(x,y,k_z)$, where $k_z=2\pi m/L_z$ for mode number $m=\{0$, $\pm1$, $\pm2$, $\ldots\}$.  The top row (a)--(c) of Fig.~\ref{fig:Bg-comp} contains plots of $E_{ES\perp1}(x,y,k_z=0)$, which is equivalent to the spatial average $\langle E_{ES\perp1}\rangle_z$.   The green arrows indicate the local direction (but \textit{not} the magnitude) of the $z$-averaged projection of $\mathbf{B}$ in the $x$--$y$ plane. Panels (a)--(c) can be interpreted as the Hall electric field, with the red regions near the separatrix branches corresponding to regions where $\mathbf{E}$ points inward (i.e., toward the exhaust).  The $z$-averaged behavior corresponds closely with what is observed in 2D simulations. In particular, there is a transition from a predominantly inward $E_{\mathrm{Hall}}$ on all four separatrix branches at low $B_g$ to a bipolar $E_{\mathrm{Hall}}$, especially on the upper-left and lower-right branches, for larger values of $B_g/B_0$. 

The second row (a')--(c') and third row (a'')--(c'') contain the magnitude and complex phase of the \textit{dominant} Fourier component with $m\ne 0$.  We identify the dominant component as the one for which $|E_{ES\perp1}(x,y,k_z)|$ has the larges \textit{global} maximum. (Because $E_{ES\perp1}(x,y,-k_z)=E_{ES\perp1}^*(x,y,k_z)$, we only have to consider positive values of $m$.)  As $B_g/B_0$ increase, the dominant value of $m$ decreases from $m=8$ to $m=6$, corresponding to an increase in the wavelength of the perturbation $\lambda_z$ from $1.25d_{i0}$ to $1.67d_{i0}$.  More significant is the transition from the perturbations being localized to the DF at $B_g/B_0=0.1$ (a') to being localized almost exclusively to the two separatrix branches with the strongly bipolar Hall $E$ signature at $B_g/B_0=1.0$ (c').  The intermediate case with $B_g/B_0=0.5$ (b') shows strong perturbations in both regions.

The complex phase (a'')--(c'') is represented by the HSV (hue, saturation, value) color as depicted in the upper-right inset in panel (c'').  Here, H is determined by the complex phase ($\tan^{-1}[Im(E)/Re(E)]$) and S is determined by the magnitude of $|E|/|E|_{\mathrm{max}}$, with V$\equiv 1$. In the inset, S decrease logarithmically by one order of magnitude from center to edge.   While the absolute phase has no significance, the phase gradient is an indicator of the local \textit{effective} $\mathbf{k}$ projected into the $x$--$y$ plane. Specifically, the normal to the bands of constant H indicate the direction of $\mathbf{k}_{\mathrm{eff}}$ and the spacing between bands of equal H indicate the effective wavelength. The phase-fronts associated with the separatrix instability show a tighter spacing in panel (c'') than in (b''), suggesting that the $x$--$y$ projection of the wavevector increases for larger values of $B_g$ as the dominant $k_z$ decreases.

\section{Separatrices as energizers}
\label{sect-energy}

\begin{figure}
\centering
   \includegraphics[width=.6\columnwidth]{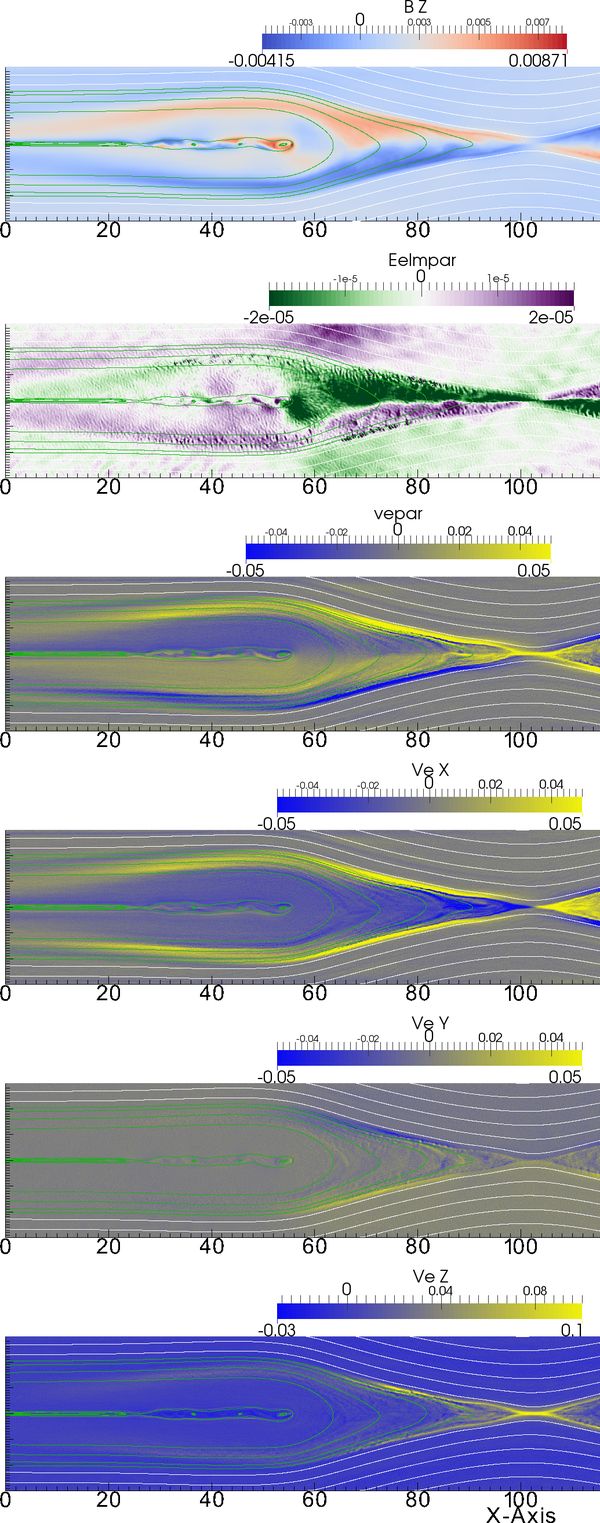}
  \caption{Run OpenBg1, time $\omega_{ci}t = 34.7$.  From top to bottom: out of plane magnetic field ($B_z$), electromagnetic component of the parallel electric field (parallel to the magnetic field), parallel electric velocity and in order the three cartesian components of the electron velocity.}
\label{sep-accel}
\end{figure}

Electrons and ions are energized by a variety of processes. A vast literature exists on the topic,  with  excellent recent reviews (see e.g. the section by Hoshino in \citet{priest-forbes} and in \citet{birn2012particle}). Here we focus on the role of the separatrix only. Two types of energization need to be distinguished at the separatrices even more so than in general: bulk acceleration and heating. 

Bulk acceleration refers to a fluid moment point of view, where the local average speed is increased. At the separatrices, the local intense electric fields lead to bulk acceleration. Figure \ref{sep-accel} reports some key fields to provide a view of the bulk flow energization. The signature of the Hall fields is associated with a parallel electric field, in tho case electromagnetic in nature.  To obtain this signature, the electrostatic component is subtracted from the total parallel electric field and the resulting electromagnetic component is reported in panel  b. The electrostatic component is obtained from the Gauss theorem and the electromagnetic part is obtained simply as a difference between the total field computed from Maxwell's equations and the electrostatic part. 

As can be seen, the parallel electric field has the same shape as the Hall magnetic field. As noted above, both can be considered part of a KAW wing that forms from the flow across the separatrices and the x-point. The KAW wing, then incorporates a self consistent parallel electric field capable of generating an electron flow that also presents the same footprint of the Hall magnetic field and of the parallel electric field. 

However, there is a second component to the electron velocity that is concentrated on a much narrower layer across the separatices propers and closer to the x-point. These narrow bands of  velocity are visible in the $z$ component and are predominant in the parallel direction. As noted in Fig.~\ref{unfrozen} and the discussion connected with it, these intense and narrow speed channels are associated also with a violation of the  frozen in condition along the separatrices.

Heating is the second type of species energization and is due to the increase of the thermal motion of the particles around their thermal speed.

\begin{figure}
  \centering
  \includegraphics[width=\columnwidth]{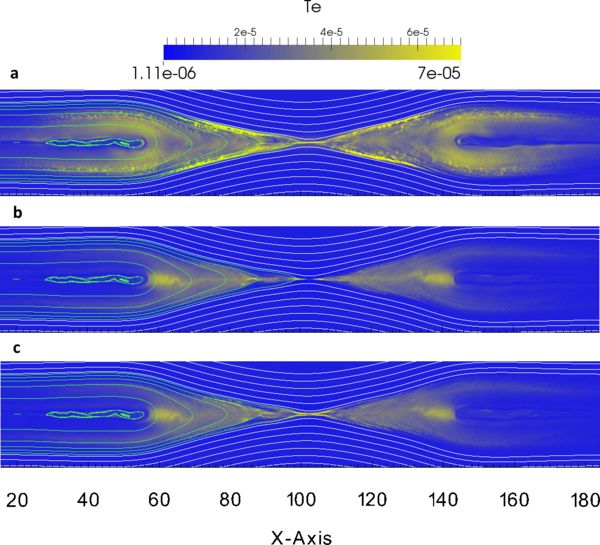}
  \caption{Run OpenBG1: Electron temperatures at time $\omega_{ci}t=34.7$ for the run described in Appendix A. Top (a): parallel, Middle (b) perpendicular in plane, Bottom (c) perpendicular out of plane. Non gyrotropy is evident only in the very close vicinity of the x-point. Anisotropy, instead, is strongest at the separatrices.  }
\label{et}
\end{figure}

\begin{figure}
  \centering
  \includegraphics[width=\columnwidth]{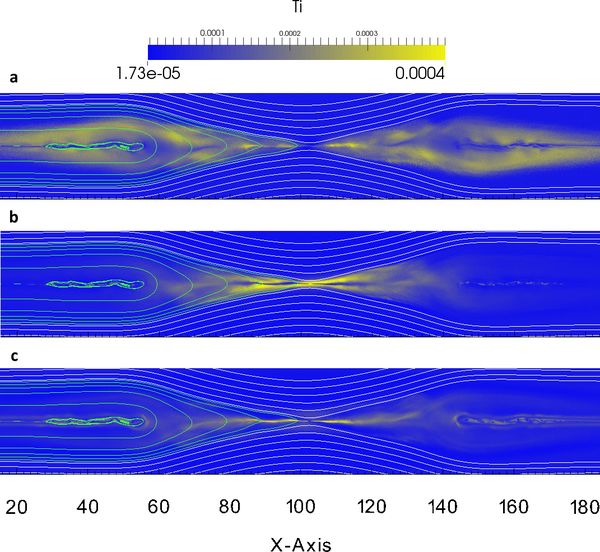}
  \caption{Run OpenBG1: Ion temperatures at time $\omega_{ci}t=34.7$ for the run described in Appendix A. Top (a): parallel, Middle (b) perpendicular in plane, Bottom (c) perpendicular out of plane. Parallel and perpendicular heating are very different,leading to anisotropy in most of the region encompassed by reconnected field lines. Non gyrotropy is also strong in the two large areas in the outflow.   }
\label{it}
\end{figure}

Figures \ref{et}-\ref{it}  report the particle temperature in the parallel and in the two perpendicular directions, respectively for electrons and ions. 
The electron heating at the separatrices is primarily in the parallel direction. The perpendicular heating and the ion heating is instead more pronounced in the reconnection exhaust~\citep{ashour2011observations,birn2012particle}.
The separatrix is the location of the interface between a region still populated by unperturbed cold particles tied to unreconnected field lines and the reconnection exhaust populated by heated particles. What happens at the separatrix to cause this transition?

A obvious consideration is that electric fields do not heat directly: their effect is to accelerate the plasma species macroscopically, leading to increased species flow speed, not directly to their thermal energy. It is useful to recall the energy balance equations. 

%

The source of thermal energy is not, as can be seen, the electric field that directly affects only the bulk energy. The source of thermal energy is the work done by the pressure tensor via the term $\bfu \cdot \nabla \cdot \tsP$, a generalisation of the famous PdV work that propelled the industrial revolution.

\begin{figure}
  \centering
  \includegraphics[width=\columnwidth]{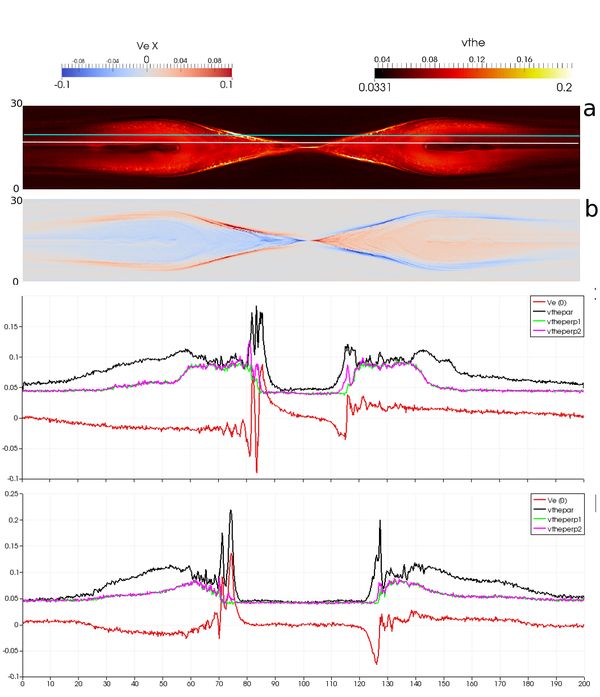}
  \caption{Run OpenBg1, time $\omega_{ci}t = 34.7$. Electron and parallel thermal and directed speed (component x). The two top panels show: a) Parallel thermal speed, $v_{th,e}/c$, b) directed speed $V_{ex}/c$. The top panel (a) indicates also two crossing lines where the speeds ($x$-component of the bulk speed and thermal speeds in the parallel and two perpendicular direction) are measured and reported on panels c and d, respectively at $y/d_i=18$ and $y/d_i=20.5$.   }
\label{combo_cross}
\end{figure}

To find the cause of electron and ion heating at the separatrices, we then need to turn our attention to the physics behind the structure of the pressure tensor and not to the electric field. As noted above, the separatrices are regions of intense flows, shears and phase space instabilities.  Figure \ref{combo_cross} reports the parallel thermal speed and the directed horizontal speed at time $\omega_{ci}t = 34.7$ in the reference run used so far. The separatrix divides the outer flow, directed towards the center, from the inner exhaust flow, directed away. At the separatrices themselves, the flow is sheared and stripes forms,  related to the instabilities  discussed in Sect.~\ref{sect-instabilities}. 

Figure \ref{combo_cross} shows the local thermal and directed speeds along two lines indicated in panel a and positioned horizontally at different distances from the neutral line. This diagnostics is conducted in an analogy with probe crossings from spacecrafts. Of course, here the data is all from the same time, at different locations, when, instead, spacecraft data records the passage of time and measures the different conditions as the plasma passes by the spacecraft (whose speed is negligible compared to that of the plasma it is immersed in). 

At all locations, the thermal speed dominates, but significant bulk flows are present, with strong local gradients. These conditions are conductive to the instabilities described above in Sect.~\ref{sect-instabilities} with important consequences to particle heating.

\begin{figure}
  \centering
  a)\\
  \includegraphics[width=\columnwidth]{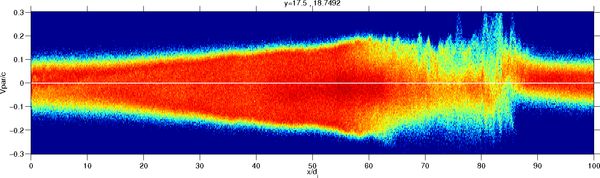}\\
  b)\\
   \includegraphics[width=\columnwidth]{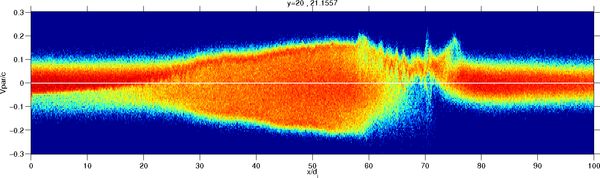}
  \caption{Run OpenBg1, time $\omega_{ci}t = 34.7$. Phase space $(x/d_i,v_{||}/c)$ along two narrow bands centred around, respectively, $y/d_i=18$ (a) and $y/d_i=20.5$ (b).  The color scale is not reported because the values are normalised to the number of particles in the box. The color scale ranges from blue to red in the conventional way, using a logarithmic scale.}
\label{phase-space}
\end{figure}

Figure \ref{phase-space} illustrates the mechanism of particle heating at the separatrices~\citep{Divin2012}. The electron thermal speed is measured as the second order moment in v, around the average speed. It is often non-trivial to "measure" the thermal spread by eye, especially because the bulk with the darker red weights much more. For this reason,  Fig.~\ref{combo_cross} reports the actual value corresponding to the phase spaces in Fig.~\ref{phase-space}.  

Three regions are evident in both crossings. The central region (from 90 to 110 on the lower crossing and from  80 to 120 in the upper crossing) is still that of the initial unperturbed plasma loaded initially in the simulation. Corresponding the same range in phase space is still made of thermal maxwellian plasma. At the other ends, leftward of the left DF (located at about $x/d_i=50$) and rightward of the right DF (located at about $x/d_i=150$), the plasma electrons are heated to about twice the initial unperturbed temperature. This is the outcome of Fermi and betatron heating associated with the DF. Several studies, experimental and theoretical,  have concentrated on the study of the electron heating at the DF~\citep{ashour2011observations,pan2012adiabatic, birn2013particle, runov2013electron}. Two  acceleration mechanisms are activated by the DF: adiabatic heating (betatron) induced by the increase in the vertical magnetic field of the DF and mirroring (Fermi). The adiabatic heating operates on the magnetised electrons. Moving from the x-point to the DF, the conservation of the first adiabatic moment, suggests that $v_{th,\perp}^2/B$ is constant and the increase in $B$ from the x-point to the DF accounts for perpendicular heating. However, electrons heated in the perpendicular direction, transform their thermal speed between  parallel and perpendicular direction as they mirror along the curved  more-dipolar lines. The DF forms a sort of radiation belt of heated particles that mirror between the northern and southern lobes. This process requires including the Earth magnetic field to be described properly, a feature absent form the preset simulation. Nevertheless, the process of reconnection has created curved magnetic field lines that increase their density at the two horizontal ends of the box, left and right above and below the central sheet. 
Additionally, the mirror motion also causes heating directly as the field lines are adveced in the exhaust and the mirroring points become closer, leading to Fermi acceleration. These two processes, adiabatic heating and Fermi acceleration~\citep{lichtenberg1980fermi}, combine to give the particle heating observed from the DF outward. 

A third region is evident in between the central unperturbed plasma and the external plasma heated by the DF. These two intermediate regions are where the separatrices passes. These two regions are between 
$x/d_i=60-90$ on the lower crossing (and symmetrically on the other side, not shown) and $x/d_i=60-85$ on the upper crossing  in Fig.~\ref{phase-space}. Here the exhaust plasma and the fresh inflow plasma come together at the separatrices, forming a layer of high shears where the streaming instabilities described in Sect.~\ref{sect-instabilities} cause a intense phase-space mixing. This leads to intense heating and the largest parallel temperatures observed. The separatrices instabilities are then not just an indicator of the separatrix region, but act to change profoundly the properties the electrons passing through. 

The phase space plots  in Fig.~\ref{phase-space} show clearly the presence of a chain of electron holes in  phase space, the non-linear consequence of the streaming instabilities~\citep{Goldman2008, Goldman2014}.

The electric field accelerates particles in the parallel direction, as discussed in the sections above, but the beams formed interact with the background plasma, with each other and with the ion neutralising background. The resulting effect is to turn the directed bulk energy of beams into thermal heating of the plasma. Electron holes caused by the beam interaction are the smoking gun that give direct evidence of the mechanism behind the acceleration observed by \citet{Divin2012}. In phase space, the particle distribution assumes the flat top formed by filling the velocity space between the interacting beams and the bulk plasma. Spacecraft observations have noted the presence of such distributions during separatrix crossings from the Cluster mission~\citep{asano2008electron,2013AGUFMSM13B2145W}.


%
%
%
%

\section{Summary}
\label{sect-conclusions}
The MMS (Magnetospeheric Multi Scale) mission is nearing its launch date and the theory community has been preparing over  the last several years to support the multidisciplinary study of reconnection. A focus of the mission is to observe the fastests and most localised features of the electron physics. We recap here our most recent results in support of the mission, focusing only on the regions near the separatrix. Other important processes develop in the inflow region~\citep{Goldman2014}, in the outflow exhaust~\cite{vapirev2013formation} and in the central region around the reconnection~\citep{goldman-review} but are not considered here. 

We focus instead on the role of the separatrices, considering several important results obtained in the recent literature. The primary novelty of the results presented here, and never published before, is the choice of a large domain that allows the separatrices to develop on larger scales than previously published. Only with the large domains and the open boundary conditions described in Appendix, the physics of the separatrices can be described clearly. 

Tables \ref{table-sect-flow}-\ref{table-sect-instab} report the key results reported in the present study relative to the flow pattern, the Hall physics and the instabilities. Table \ref{table-sect-energy} reports the new results and the new physics of electron and ion energisation reported in the present work for the first time. The study of energy balance is the most recent development in our investigations of reconnection. Previously, we have considered the energetics of the outflow~\citep{lapenta2014electromagnetic} but the present paper is the first focusing on the energetic exchange at the separatrices. 

The Table \ref{table-sect-energy} identifies a key role of the separatrix instabilities: their role in increasing the electron parallel temperature via the streaming instability the phase mix different electron populations leading to larger second moments of the distribution. These are hot electrons whose distribution is very far from Maxwellian and the concept of temperature must be considered in its kinetic sense, because the local thermodynamic equilibrium is certainly not satisfied. 

\begin{table}
\begin{tabular}{|p{.8\textwidth}|p{.2\textwidth}|} \hline
Key result & Key Figure\\
  \hline
  {\it Flow Deflection at the separatrix}: both species receive a great localised deflection at the separatrix turning from a mostly vertical flow to a motley horizontal flow & Fig.~\ref{eflow}-\ref{iflow} \\
  {\it Electron parallel Flow}: near the separatrices the electrons acquire a strong parallel flow  &  Fig.~\ref{eflow}\\
   {\it Electron not frozen-in}: at a layer across the separatrice the electrons are subjected to very localised forces and the drift approximation fails, with the frozen in condition being violated, as strongly as in the central diffusion region  &  Fig.~\ref{unfrozen}\\
 {\it Effect of a guide field}: the flow pattern near the separatrices changes radically wit the presence of guide fields. For significant guide fields the parallel  electron flow transitions from being always directed towards the x-point to be antisymmetric: on two separatrices it remains towards the x-point but on two others it is away from it. Both species develop a vortical flow  &  Fig.~\ref{eflowbg}-\ref{iflowbg}\\
\end{tabular}
\caption{Key points of Section~\ref{sect-flow}:  Separatrices as flow gates.}
\label{table-sect-flow}
\end{table}

\begin{table}
\begin{tabular}{|p{.8\textwidth}|p{.2\textwidth}|} \hline
Key result & Key Figure\\
  \hline
  {\it Kinetic \alfven Wave wing}:  the  Hall term term and of the divergence of the pressure tensor in the generalised Ohm's law causes a distintive signature interpreted as a kinetic \alfven wave wing  with a quadrupolar out of plane magnetic field and a in plane current pattern and net density perturbation with an associated in plane electrostatic field  & Fig.~\ref{hall_signatures} \\
  {\it Electrostatic Nature of the Hall electric Field}: the strongest electric field in the system is not the out of plane electromagnetic reconnection electric field but the in plane perpendicular field formed across the separatrices. This field is electrostatic in nature.  &  Fig.~\ref{hallE}\\
   {\it Embedded instabilities}: the Hall physics signatures are accompanied by clear signs of the presence of instabilities caused by the strong parallel flows and the strong gradients present at the separatrices  &  Fig.~\ref{other_signatures}\\
\end{tabular}
\label{table-sect-hall}
\caption{Key points of Section~\ref{sect-hall}:  Signatures of the Hall physics.}
\end{table}

\begin{table}
\begin{tabular}{|p{.8\textwidth}|p{.2\textwidth}|} \hline
Key result & Key Figure\\
  \hline
  {\it Streaming instabilities}:  Streaming instabilities are caused by the presence of relative drifts between different components of the electron distribution (two-stream instability) at the separatrices and between the electron and ion distribution  (Buneman instability) & Fig.~\ref{fig:Fig1} \\
  {\it Shear instabilities}: a layer across the separatrices has a very sheared electron velocity unstable to Kelvin-Helmholtz-type instabilities. The region presents also strong density gradients subject to dirft-type instabilities. These two sources of free energy lead to instabilities showing in 2D simulations on selected planar orientations and with multiple signatures in fully 3D simulations   &  Fig.~\ref{fig:Fig6}\\
   {\it 3D Correlations}: all instabilities possible in the separatrix layer develop in fully 3D simulations and a correlation emerges between the different instabilities so that electron holes due to parallel instabilities appear to  correlate with the striations  aused by perpendicular instabilities &  Fig.~\ref{fig:kb}\\
   {\it Effect of Guide Field}: the presence of a guide field drastically affects which instabilities become dominant. At low guide fields, the instabilities at the depolarisation fronts are more important, but at higher guide fields the separatrix instabilities take over.& Fig.~\ref{fig:Bg-comp}
\end{tabular}
\label{table-sect-instab}
\caption{Key points of Sections~\ref{sect-instabilities}-\ref{sec:3dlociofinstability}:  Instabilities at the separatrices in 2D and 3D simulations.}
\end{table}

\begin{table}
\begin{tabular}{|p{.8\textwidth}|p{.2\textwidth}|} \hline
Key result & Key Figure\\
  \hline
  {\it Bulk energy}:  The parallel electromagnetic component of the electric field structure caused but the kinetic \alfven wave wing leads to a large increase of the electron bulk energy at the separatrices. This field is obscured by the stronger electrostatic in-plane field (both parallel and perpendicular). But when isolated its role in accelerating the electrons becomes clear. & Fig.~\ref{sep-accel} \\
  {\it Species heating}: two mechanisms share the responsibility of heating both electrons and ions in the exhaust: Fermi acceleration and adiabatic heating. But for the electrons the separatrices become a dominant contribution in the direction parallel to the magnetic field   &  Fig.~\ref{et}-\ref{it}\\
   {\it Parallel heating mechanism}: the electron heating at the separatrices in the parallel direction is linked to the same regions where the frozen-in condition becomes violated at the separatrices. There the phase space deviates drastically from the distributions seen in the inflow region (Maxwellian) and in the outflow (flat-top). The electron phase space is dominated by the electron holes produces by the streaming instabilities  that convert bulk energy into thermal energy and produce the largest electron temperatures observed in the simulation. &  Fig.~\ref{phase-space}\\
\end{tabular}
\label{table-sect-energy}
\caption{Key points of Sect.~\ref{sect-energy}:  Separatrices as energizers.}
\end{table}


\bibliographystyle{jpp}

\bibliography{myabbrev,bibliografie}

\appendix
\section{Simulation approach}

The study presented here is based on the massively parallel fully kinetic code iPIC3D, extensively used in previous studies in 2D~\citep{iPIC3D,Lapenta2010,lapenta2012particle,lapenta2011bipolar} and in 3D~\citep{markidis2012three,vapirev2013formation}.The key peculiarity of iPic3D is to use an implicit time discretization that ensures an increased stability~\citep{brackbill-forslund} and better energy conservation~\citep{markidis2011energy, lapenta2011particle}. This feature allows thre user to select the desired resolution actually needed for the study without being forced to excessive over-resolution by the constraint for stability and energy conservation typical of explicit discretizations~\citep{lapenta2012particle, lapenta2013space}. In practice, this benefit allows the user to consider larger systems and more realisitc parameters, including the physical value of the mas ration for hydrogen~\citep{Lapenta2010} and oxygen~\citep{markidis2011kinetic}. In the present study, the need for the implicit method is caused by the desire to study the separatrices over large distances, requiring a relatively large box.

We consider an initial Harris equilibrium~\citep{Harris1962}:
\begin{equation}\label{initial}
\begin{array}{c}
 B_x=B_0 \tanh(y/L)\\
   \\
 n=n_b+n_0 {\rm sech}^2(y/L)
\end{array}
\end{equation}
with varying guide fields $B_z= B_g$. iPic3D does not use the GSM coordinate system. The coordinates are chosen as: $x$ along the sheared component of the magnetic field (Earth-Sun direction in the Earth magnetosphere), $y$ in the direction of the gradients (north-south in the magnetosphere) and $z$ along the current (dawn-dusk in the magnetotail).

The plasma of the Harris equilibrium is initially Maxwellian with a uniform drift that is prescribed by the force balance. A uniform background  is added in the form of a non-drifting Maxwellian at the same temperature of the main Harris plasma. This choice is not dictated by any stability of the numerical solution, as the implicit PIC method employed here has no such limits (unlike the explicit PIC approach)~\cite{birdsall-langdon}.The background is added to mimic the observed properties of satellite crossing where the current layers have a peaked density in the center and fall off to a smaller but still respectable value.

In the present study, reconnection is followed from its (artifical) onset caused by a localized X-point  perturbation described by the vector potential:
 \begin{equation}\label{perturb}
\delta A_z=A_{z0} \cos(2\pi x/L_\Delta)cos(\pi y/L_\Delta)e^{-(x^2+y^2)/\sigma^2},
 \end{equation}
with $L_\Delta=10 \sigma$ and $\sigma=d_i/2$, where $d_i$ is the ion skin depth.

The chosen physical parameters for the present study are consistent with the observed properties of the geomagnetic tail:  electron thermal velocity $v_{the}/c=.045$,  ratio of electron and ion temperatures $T_i/T_e=5$,   current sheet thickness $L/d_i=0.5$ and  background density $n_b/n_0=.1$. A reduced  mass ratio, $m_i/m_e=256$ is used to reduce the cost of the simulations (especially for the 3D case). These choices correspond also to a realistic Alfv\'en speed $c/v_a = 300$~\citep{Lapenta2010}.These ion scales are computed in the Harris density $n_0$.

We report on two types of simulations, 2D and 3D. All the 2D simulations use a computational box with $L_x/d_i=200$ and $L_y/d_i=30$ resolved with a grid of $2560x384$ cells and a time step of $\omega_{pi}\Delta t=0.125$.  All the simulations use 196 particles per cell per species, with two ion and two electron species (one for the Harris component and one for the background). 

All 3D simulations use a computational box $L_x/d_i=40$, $L_y/d_i=15$, $L_y/d_i=10$ resolved with a grid of $512x192x128$ cells and a time step of $\omega_{pi}\Delta t=0.125$ and 125 particles per species, with two ion and two electron species (one for the Harris component and one for the background).

At $t=0$ all the cells have the same number of particles whose statistical weight is chosen according to the local density which varies significantly due to the Harris equilibrium. Periodic and open boundary conditions are used in $x$ and open or  perfect conductor conditions ($\bfE_t=0$ and $\delta \bfB_n=0$) in $y$. The $z$ direction is assumed to be periodic. In 2D, just one cell is used in $z$ and this direciton becomes ignorable, but all fields and the velocities of the particles have also the $z$ component making the 2D simulation in reality 2D3V.

We have conducted several simulations, conventionally in the text tehy are indicated as either Open or Periodic, followed by BG and the indication of the first decimal of the guide field. Table~\ref{runs2d} reports the 2D cases completed and  Table~\ref{runs3d} reports the 3D cases.

\begin{table}
\begin{center}
\begin{tabular}{|c|c|c|c|}\hline \hline
Name & Guide Field $B_g/B_0$ & BC in $x$ & BC in $y$ \\ \hline
PeriodicBG0 &  0 & Periodic & Conductor\\
PeriodicBG1 &  0.1 & Periodic  & Conductor \\
OpenBG1 &  0.1 &  Open & Open\\
PeriodicBG3 &  $1/3$ & Periodic & Conductor \\
PeriodicBG5 &  $1/2$  & Periodic & Conductor\\
PeriodicBG10 &  1 & Periodic & Conductor\\
OpenBG100 &  10 & Open & Open  \\ \hline \hline
\end{tabular}
\end{center}
 \caption{2D Simulations. Cases considered and available from the MMS-IDS server of the Univerity of Colorado}
\label{runs2d}
\end{table}

\begin{table}
\begin{center}
\begin{tabular}{|c|c|c|c|} \hline \hline
Name & Guide Field $B_g/B_0$ & BC in $x$ & BC in $y$ \\ \hline
3DPeriodicBG0 &  0 & Periodic & Conductor\\
3DPeriodicBG1 &  0.1 & Periodic  & Conductor\\
3DOpenBG1 &  0.1 &  Open & Open\\
3DPeriodicBG2.5 &  0.25 &  Periodic & Conductor\\
3DPeriodicBG3.3 &  0.33 & Periodic & Conductor \\
3DPeriodicBG5 &  0.50  & Periodic & Conductor\\
3DPeriodicBG7.5 &  0.75 & Periodic & Conductor\\
3DPeriodicBG10 &  1 & Periodic & Conductor \\ \hline \hline
\end{tabular}
\end{center}
 \caption{3D Simulations. Cases considered and available from the MMS-IDS server of the Univerity of Colorado. All boundary conditions in $z$ are always periodic.}
\label{runs3d}
\end{table}

\end{document}